\documentclass[twocolumn]{aastex631}

\usepackage{booktabs}
\usepackage{textcomp}
\usepackage{tipa}

\newcommand{\plot}[1]{\textit{#1}:}

\newcommand{\rh}{$r_{\text{h}}$}
\newcommand{\rhm}{r_{\text{h}}}

\newcommand{\vh}{$v_{\odot}$}
\newcommand{\Msun}{$\textup{M}_\odot$}
\newcommand{\Lsun}{$\textup{L}_\odot$}
\newcommand{\sqdeg}{deg$^2$}
\newcommand{\kms}{km\,s$^{-1}$}
\newcommand{\met}{[Fe/H]}

\shorttitle{Ursa Major III/UNIONS 1}
\shortauthors{Smith et al.}
\graphicspath{{./}{figures/}}
\begin{document}

\title{The discovery of the faintest known Milky Way satellite using UNIONS}

\correspondingauthor{Simon E.\,T. Smith}
\email{simonsmith@uvic.ca}

\author[0000-0002-6946-8280]{Simon E.\,T. Smith}
\affiliation{Department of Physics and Astronomy, University of Victoria, Victoria, BC, V8P 1A1, Canada}

\author{William Cerny}
\affiliation{Department of Astronomy, Yale University, New Haven, CT 06520, USA}

\author{Christian R. Hayes}
\affiliation{NRC Herzberg Astronomy and Astrophysics, 5071 West Saanich Road, Victoria, BC, V9E 2E7, Canada}

\author{Federico Sestito}
\affiliation{Department of Physics and Astronomy, University of Victoria, Victoria, BC, V8P 1A1, Canada}

\author{Jaclyn Jensen}
\affiliation{Department of Physics and Astronomy, University of Victoria, Victoria, BC, V8P 1A1, Canada}

\author{Alan W. McConnachie}
\affiliation{NRC Herzberg Astronomy and Astrophysics, 5071 West Saanich Road, Victoria, BC, V9E 2E7, Canada}
\affiliation{Department of Physics and Astronomy, University of Victoria, Victoria, BC, V8P 1A1, Canada}

\author{Marla Geha}
\affiliation{Department of Astronomy, Yale University, New Haven, CT 06520, USA}

\author{Julio Navarro}
\affiliation{Department of Physics and Astronomy, University of Victoria, Victoria, BC, V8P 1A1, Canada}

\author{Ting S. Li}
\affiliation{Department of Astronomy and Astrophysics, University of Toronto, 50 St. George Street, Toronto ON, M5S 3H4, Canada}

\author{Jean-Charles Cuillandre}
\affiliation{AIM, CEA, CNRS, Universit\'e Paris-Saclay, Universit\'e Paris, F-91191 Gif-sur-Yvette, France}

\author{Rapha\"el Errani}
\affiliation{Universit\'e de Strasbourg, CNRS, Observatoire Astronomique de Strasbourg, UMR 7550, F-67000 Strasbourg, France}

\author{Ken Chambers}
\affiliation{Institute for Astronomy, University of Hawaii, 2680 Woodlawn Drive, Honolulu HI 96822}

\author{Stephen Gwyn}
\affiliation{NRC Herzberg Astronomy and Astrophysics, 5071 West Saanich Road, Victoria, BC, V9E 2E7, Canada}

\author{Francois Hammer}
\affiliation{GEPI, Observatoire de Paris, University\'e PSL, CNRS, Place Jules Janssen F-92195, Meudon, France}

\author{Michael J. Hudson}
\affiliation{Department of Physics and Astronomy, University of Waterloo, 200 University Ave W, Waterloo, ON N2L 3G1, Canada}
\affiliation{Waterloo Centre for Astrophysics, University of Waterloo, 200 University Ave W, Waterloo, ON N2L 3G1, Canada}
\affiliation{Perimeter Institute for Theoretical Physics, 31 Caroline St. North, Waterloo, ON N2L 2Y5, Canada}

\author{Eugene Magnier}
\affiliation{Institute for Astronomy, University of Hawaii, 2680 Woodlawn Drive, Honolulu HI 96822}

\author{Nicolas Martin}
\affiliation{Universit\'e de Strasbourg, CNRS, Observatoire Astronomique de Strasbourg, UMR 7550, F-67000 Strasbourg, France}
\affiliation{Max-Planck-Institut f\"ure Astronomie, K\"onigstuhl 17, D-69117, Heidelberg, Germany}



\begin{abstract}

We present the discovery of Ursa Major III/UNIONS 1, the least luminous known satellite of the Milky Way, which is estimated to have an absolute V-band magnitude of $+2.2^{+0.4}_{-0.3}$\,mag, equivalent to a total stellar mass of 16$^{+6}_{-5}$\,\Msun. 
Ursa Major III/UNIONS 1 was uncovered in the deep, wide-field Ultraviolet Near Infrared Optical Northern Survey (UNIONS) and is consistent with an old ($\tau > 11$\,Gyr), metal-poor ([Fe/H] $\sim -2.2$) stellar population at a heliocentric distance of $\sim$\,10\,kpc. Despite being compact ($r_{\text{h}} = 3$\,$\pm$\,1\,pc) and composed of so few stars, we confirm the reality of Ursa Major III/UNIONS 1 with Keck II/DEIMOS follow-up spectroscopy and identify 11 radial velocity members, 8 of which have full astrometric data from {\it Gaia} and are co-moving based on their proper motions. Based on these 11 radial velocity members, we derive an intrinsic velocity dispersion of $3.7^{+1.4}_{-1.0}$\,\kms\ but some caveats preclude this value from being interpreted as a direct indicator of the underlying gravitational potential at this time. 
Primarily, the exclusion of the largest velocity outlier from the member list drops the velocity dispersion to $1.9^{+1.4}_{-1.1}$\,\kms, and the subsequent removal of an additional outlier star produces an unresolved velocity dispersion. 
While the presence of binary stars may be inflating the measurement, the possibility of a significant velocity dispersion makes Ursa Major III/UNIONS 1 a high priority candidate for multi-epoch spectroscopic follow-ups to deduce to true nature of this incredibly faint satellite.

\end{abstract}



\section{Introduction} \label{sec:intro}

%Wide field surveys have revealed a huge amount of substructure in the Milky Way Halo (list a few notable dwarfs, streams, faint star clusters found pre Gaia). The addition of Gaia has allowed for even more identification due to proper motions confirming things.
%Small blurb about dwarf galaxies and globular clusters as traces of Milky Way evolutionary history, dark matter.
%Faint star clusters are an ambiguous class of object and exist moreso as a collection of systems that do not fit the typical traits of globular clusters or UFDs. 

Wide-field, digital, photometric surveys have revealed a rich landscape of substructure in the Milky Way halo since their inception in the early 2000s. The Sloan Digital Sky Survey \citep[SDSS;][]{Abazajian09} made an immense impact, brokering the discoveries of many faint Milky Way dwarf galaxy satellites \citep[e.g.][]{Willman05-w1, Willman05-uma, Zucker06, Belokurov06-boo, Belokurov07-quin, Belokurov09, Belokurov10}.
%and the wispy trails of stellar streams \citep{Belokurov06-field, Belokurov07-orphan}. 
Another waterfall of Milky Way dwarf galaxy discoveries \citep[e.g.][]{Laevens15-three, Laevens15-one, Bechtol15, Drlica15, Koposov15-beasts, Homma16, Homma18} came with the assembly of photometric catalogs such as Pan-STARRS 3pi \citep{Chambers16}, the Dark Energy Survey \citep[DES;][]{Abbot18}, and the HyperSuprime-Cam Subaru Strategic Program \citep[HSC-SSP;][]{Aihara18-overview, Aihara18-Y1}. 

Ongoing surveys, such as the Dark Energy Local Volume Explorer \citep[DELVE;][]{Drlica21, Drlica22} and the Ultraviolet Near-Infrared Optical Northern Survey \citep[UNIONS;][]{Ibata17}, have continued to bolster the known Milky Way satellite dwarf galaxy population \citep{Mau20, Cerny21, Cerny23-peg, Cerny23-6, Smith22}. The {\it Gaia} space telescope \citep{Gaia16, Gaia21} has also been revolutionary, with its proper motion measurements providing additional constraints for characterising and discovering both dwarf galaxies \citep[e.g.][]{Torrealba19-ant2, Pace19, McVenn20a, McVenn20b} and globular clusters \citep[e.g][]{Torrealba19-9, Pace23}. 

Classical globular clusters are typically bright and compact stellar systems, while dwarf galaxies are orders of magnitude more diffuse than globular clusters at similar magnitudes and cover a much broader range in characteristic size. \citet{Willman12} proposed that the key physical distinction between these two types of systems is that the dynamics of dwarf galaxies cannot be explained through a combination of baryonic processes and Newton’s laws, while globular clusters can be explained in such a way. Therefore, in the framework of $\Lambda$CDM cosmology, dwarf galaxies are thought to lie at the center of their own dark matter halos. The faintest known dwarf galaxies \citep[sometimes called ultra-faint dwarf galaxies or UFDs;][]{Simon19} are observed to have dynamical masses (measured from stellar kinematics) many orders of magnitude larger than the mass implied by total luminosity (dynamical mass-to-light (M/L) ratios $\sim$ 10$^3$\,\Msun/\Lsun). 
%Such dark matter dominated systems have proven to be powerful probes of star formation \citep[e.g. gas retention;][]{Bovill09}, quenching via reionization \citep[e.g.][]{Bullock00}, and chemical enrichment \citep[e.g.][]{Ji16-tuc2, Ji16-ret2, Ji23, Hayes23} paradigms as well as theories of dark matter and structure evolution both on a cosmological scale \citep[e.g.][]{Lovell12, Bullock17} and in the Milky Way environment \citep[e.g.][]{Wheeler15, Applebaum21}.
Dynamical analysis of globular clusters, on the other hand, shows that they do not have appreciable amounts of dark matter and are comprised solely of baryonic matter.

\citet{Willman12} also suggested that a significant dispersion in the distribution of stellar metallicities could be used as a proxy for the presence of a dark matter halo.
It is argued that the shallow potential wells of globular clusters are unable to retain the products of stellar feedback and thus form a stellar population of a single metallicity. In contrast, it is argued that dwarf galaxies can retain gas and have prolonged star formation histories, leading to self-enrichment. Significant metallicity dispersions have often been used to distinguish between dwarf galaxies and globular clusters in this way \citep[e.g. ][]{Leaman12, Kirby13, Li22}.

In parallel with the increase of dwarf galaxy discoveries, a number of faint Milky Way satellites of ambiguous nature have also been unearthed.
These faint systems are typically small in physical extent (half-light radius, $r_{\text{h}} \lesssim 15$\,pc), well within the virial radius of the Milky Way (heliocentric distance, $D_\odot \lesssim 100$\,kpc), and faint (absolute $V$-band magnitude, $M_V \gtrsim -3$\,mag)\footnote{See Appendix \ref{ap:sats} for a full list of references}. Beyond these general observations, these tiny Milky Way satellites are still poorly understood for two reasons: (1) In these observed properties, they lie at the interface of dwarf galaxies and globular clusters, and (2) their internal dynamics and chemical properties are not well studied en masse. 

Diagnostics such as size \citep[e.g.][]{Balbinot13, Conn18}, stellar mass segregation \citep[e.g.][]{Koposov07-fsc, Kim15-K2}, and comparison to the dwarf galaxy stellar mass-metallicity relation \citep[e.g.][]{Jerjen18} have been used to argue that some of these systems are more likely to be ultra faint star clusters (i.e. lacking dark matter). However, neither the presence nor lack of a dark matter halo has been demonstrated conclusively for any one of these system.

Dynamically confirming the nature of any one of these faint, ambiguous satellites could extend either the globular cluster or dwarf galaxy luminosity functions by up to a few orders of magnitude, and could extend the dwarf galaxy scale-length function by up to a factor of 10.
Globular clusters are valuable for studying the evolution of the interstellar medium and stellar populations over cosmic time \citep[e.g.][]{Krumholz19, Adamo20} while dwarf galaxies have proven to be powerful probes of star formation \citep[e.g.][]{Bovill09}, chemical enrichment \citep[e.g.][]{Ji16-tuc2, Ji16-ret2, Ji23, Hayes23}, and the nature of dark matter \citep[e.g.][]{Lovell12, Wheeler15, Bullock17, Applebaum21}. The faintest and smallest dwarf MW satellites are particularly constraining. Their total number may be used to constrain alternative  models, such as ``warm'' dark matter \citep{Lovell12}, or ``fuzzy'' dark matter \citep{Nadler21}. In addition, their characteristic densities place strong constraints on self-interacting dark matter models \citep{Errani22, Silverman23}. Additional studies of these faint, ambiguous systems, including radial velocities and metallicity measurements, will be needed to further understand individual satellites as well as how the characteristics of globular cluster and dwarf galaxy populations extend to such faint magnitudes and parsec-length scales.

In this paper, we detail the discovery and characterization of Ursa Major III/UNIONS 1, the least luminous Milky Way satellite detected to date. Line-of-sight velocities of candidate member stars obtained through follow-up spectroscopic observations may imply a significant radial velocity dispersion, but repeat radial velocity measurements are be needed to conclusively demonstrate whether dark matter is present in this system. We refer to this system as Ursa Major III/UNIONS 1 as its identity as a dwarf galaxy or star cluster is not clear at this time.
%We do not find conclusive evidence to either support or contradict the presence of a surrounding dark matter halo, meaning that the classification of Ursa Major III/UNIONS 1 remains ambiguous.
%Although a conclusive classification is not reached, the balance of evidence at this time suggests that Ursa Major III/UNIONS 1 is an extremely faint star cluster.
In Section \ref{sec:methods} we summarize the discovery dataset, detection of the system, and follow-up spectroscopy. In Section \ref{sec:results}, we characterize the structural parameters of Ursa Major III/UNIONS 1, as well as its distance, luminosity, dynamics, and orbit. Finally, in Section \ref{sec:disc}, we discuss the classification of Ursa Major III/UNIONS 1, and summarize our results.

\section{Data and Detection} \label{sec:methods}

\subsection{UNIONS}\label{subsec:data}

Ursa Major II/UNIONS 1 (UMa3/U1) was discovered during an ongoing search for faint Local Group systems in the deep wide-field Ultraviolet Near Infrared Optical Northern Survey (UNIONS). UNIONS is a consortium of Hawaii-based surveys working in conjunction to image a vast swath of the northern skies in the $ugriz$ photometric bands. 
Four distinct surveys are contributing independent imaging: the Canada-France Imaging Survey (CFIS) at the Canada-France-Hawaii Telescope (CFHT) is targeting deep $u$ and $r$ photometry, Pan-STARRS is obtaining deep $i$ and moderately deep $z$ observations, the Wide Imaging with Subaru HyperSuprime-Cam of the Euclid Sky (WISHES) program is acquiring deep $z$ at the Subaru telescope, and the Waterloo-Hawaii IfA G-band Survey (WHIGS) is responsible for deep $g$ imaging, also with Subaru. Together, these surveys are covering 5000\,\sqdeg\ at declinations of $\delta > 30 \deg$ and Galactic latitudes of $|b| > 30 \deg$. UNIONS was in part brought together to support the {\it Euclid} space mission, providing robust ground-based $ugriz$ photometry necessary for photometric redshifts that will be the main pillar of {\it Euclid}'s science operations. However, UNIONS is a separate survey whose aim is to maximize the science returns of this powerful, deep, wide-field photometric dataset. UNIONS aims to deliver 5$\sigma$ point source depths of 24.3, 25.2, 24.9, 24.3, 24.1\,mag in $ugriz$ which is roughly equivalent to the first year of observations expected from the Legacy Survey of Space and Time (LSST) at the Vera C. Rubin Observatory, making UNIONS the benchmark photometric survey in the northern skies for the coming decade.

This work only utilizes the CFIS-r and Pan-STARRS-i band catalogs. The median 5$\sigma$ point source depth in a 2 arcsecond (\arcsec) aperture is 24.9\,mag in $r$ and 24.0\,mag in $i$, although the Pan-STARRS-i depth will increase over time owing to the scanning-type observing strategy employed. Currently, the area common to the two datasets spans more than 3500\,\sqdeg\ across both the North Galactic Cap (NGC) and South Galactic Cap (SGC). These catalogs were cross-matched with a matching tolerance of 0.5\arcsec, though sources typically matched to better than 0.13\arcsec.

The median image quality of the CFIS-r observations is an outstanding 0.69\arcsec, allowing us to perform star galaxy separation morphologically. We correct for galactic extinction using $E(B-V)$ values from \citep{Schlegel98} assuming the conversion factors given by \citep{Schlafly11} for a reddening parameter of $R_V = 3.1$. The CFHT $r$-band filter is not present in the \citet{Schlafly11} list, so we adopt the conversion factor for Dark Energy Camera (DECam) $r$-band filter \citep{Flaugher15} given that their full-width at half-maximum (FWHM) are identical, and that the DECam-r centroid is shifted redwards from CFHT-r by only 2\,nm. For the remainder of the manuscript, $r$ will refer to CFIS-r and $i$ will refer to Pan-STARRS-i unless otherwise specified.

%Later on in this analysis, we also utilize data from the Third Data Release (DR3) from {\it Gaia} \citep{Gaia16, Gaia21} and we extinction-corrected the photometry following \citet{Gaia18}.

\begin{figure*}
    \centering
    \includegraphics[width=1\linewidth]{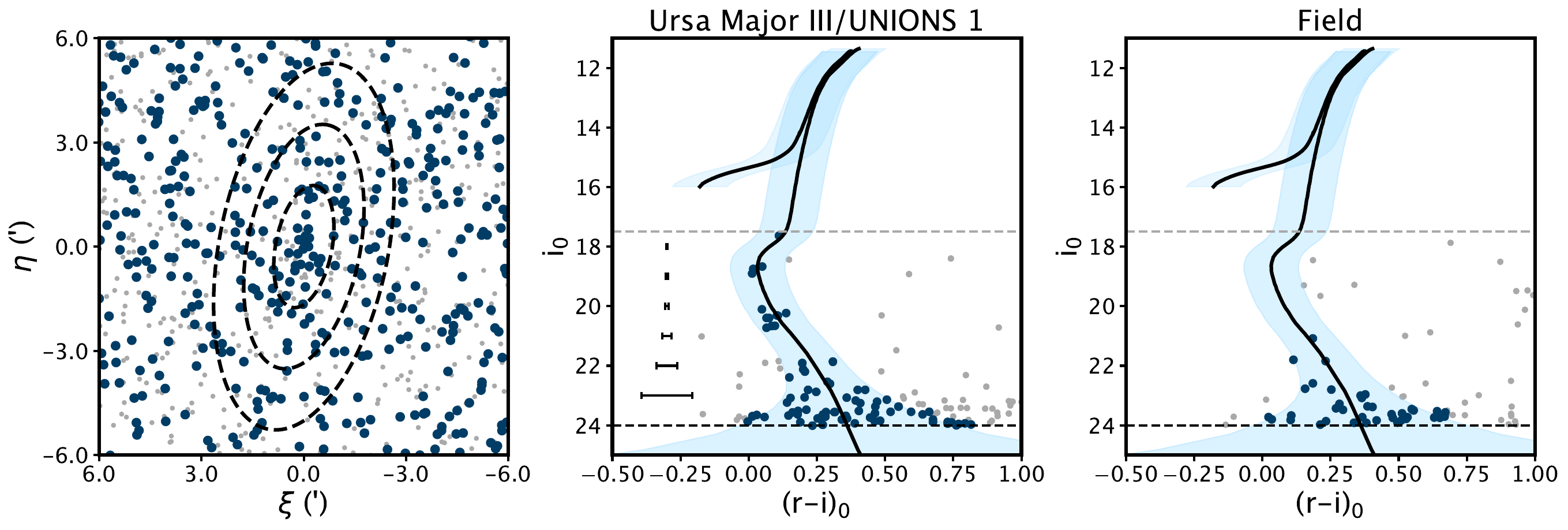}
    \caption{Detection Plot for UMa3/U1. 
    \plot{Left} Tangent plane-projected sky positions of all stars within a 12\arcmin\,$\times$ 12\,\arcmin\ region around the overdensity. Isochrone-selected points are colored blue while unselected stars are grey. The dashed black ellipses have semi-major axes of 2\,$\times$, 4\,$\times$, and 6\,$\times$ the half-light radius (\rh), where \rh\ is determined by the MCMC structural-parameter estimation routine described in Section \ref{subsec:structural}.
    \plot{Center} Color-magnitude diagram (CMD) of extinction-corrected $r, i$ photometry for all stars within 4\,$\times$\,\rh\ ellipse. An old (12\,Gyr), metal-poor (\met\,=\,$-2.2$) isochrone shifted to a distance of 10\,kpc is overlayed (black), along with the matched-filter selection region (light blue). Sources are colored as in the left-hand panel. The median photometric uncertainties as a function of $i$ magnitude are shown in black on the left side of the CMD.
    The median 5$\sigma$ point source depth of $i$ (24\,mag) is shown as a black dashed line while the approximate saturation limit (17.5\,mag) is shown as a grey dashed line.
    \plot{Right} Same as central panel, for all stars in an outer annulus with area equal to the area enclosed by the 4\rh\ ellipse. Nominally, these sources should be comprised of Milky Way halo stars and incorrectly classified faint background galaxies, which demonstrates the level of contamination on the UMa3/U1 CMD.
    }
    \label{fig:cmd-detect}
\end{figure*}

\subsection{Detection Method}\label{subsec:detect}

UMa3/U1 was discovered as a spatially resolved overdensity of stars using a matched-filter approach. Variations on this general methodology have proven to be efficient and productive in discovering faint dwarf galaxies and faint star clusters in wide-field surveys \citep[e.g.][]{Koposov07-fsc, Walsh09, Bechtol15, Drlica15, Koposov15-beasts, Homma18}. 

Our particular implementation of the algorithm will be described here in brief, though we refer the reader to \citet{Smith22} for a more detailed overview of the search method. A matched-filter seeks to isolate stars that belong to some particular stellar population, in order to create contrast in stellar density between the member stars of a putative satellite and the Milky Way stellar background. We start by selecting all stars in the UNIONS footprint that are consistent with a 12\,Gyr, \met$=-2$ PARSEC isochrone \citep{Bressan12}, constructed from the CFHT-r and PanSTARRS-i bands, with the isochrone shifted to a test distance. Stars are then tangent plane projected, binned into 0.5\arcmin\,$\times$\,0.5\arcmin\ pixels, and smoothed with Gaussian kernels with FHWMs of 1.2, 2.4, and 4.8\arcmin. The broad-scale mean and variance are then subtracted and divided out from the smoothed maps, respectively, to find the significance of each pixel with respect to the Milky Way background. All peaks in the significance map are then recorded for future examination, and this process is repeated for a series of heliocentric distances spaced roughly logarithmically from 10\,kpc to 1\,Mpc.

Our match-filter algorithm has been successful in detecting previously known dwarf galaxies. It should be noted that dwarf galaxy detections are the main focus of this search, but extra-galactic star clusters are also typically old and metal-poor, so the matched-filter ought to pick them up as well. We have used the known dwarf galaxy population to do a first-pass assessment of the algorithm's efficiency. Within 1\,Mpc, we have recovered, with high statistical significance, all known Local Group dwarf galaxies (including M31 dwarfs) that had been found in shallower surveys with matched-filter methods. 
%Only Bo\"otes IV (cite) and Pegasus V (cite) have eluded detection, though Bo\"otes IV was found in the deeper HSC survey and Pegasus V, an M31 satellite, was found in a dedicated visual search through the DESI Legacy Imaging Surveys. 
We also detect other galaxies in the Local Universe out to 2\,Mpc, and several Milky Way globular clusters.

%We have since moved onto our lists of significance peaks to investigate previously unknown groups of spatially clustered stars. 
UMa3/U1 is one of the most prominent candidates without a previous association produced from our search, on par with the detection significance of some known UFDs in the UNIONS sky. UMa3/U1 is most prominently detected in the significance map produced by the 1.2\arcmin\ smoothed, 10\,kpc iteration at a statistical significance of 3.7$\sigma$ above the background. For reference, Draco II is an ultra-faint satellite of the Milky Way whose true nature is unknown, but it has been estimated to be $\sim 20$\,kpc distant, and is detected at 2.8$\sigma$ above the background in our own search. 
Figure \ref{fig:cmd-detect} shows a color-magnitude diagram (CMD) using $r, i$ photometry, demonstrating the detection of stars about UMa3/U1 that meet the matched-filter selection criteria. 
We also show an equivalent CMD of a reference field, which is an equal-area elliptical annulus, to indicate the expected level of source contamination in the detection.

\subsection{Keck/DEIMOS Spectroscopy}\label{subsec:keck}

%Write up description of observations. Use other Keck/DEIMOS papers as inspiration and check in with Will about a few other specifics. I think he passed along the exposure times. Can include a table of radial velocity members with S/N, PSAT, rvs, fe/h, pm...

%Target selection

%Observations

%Reduction
%Describe reduction methods via 'Geha et al, in prep.' and ask for more details as needed.

We obtained spectroscopic data for 59 stars towards UMa3/U1 using the DEIMOS spectrograph \citep{Faber03} on the Keck II 10-m telescope. Data were taken on April 23rd, 2023 using a single mutlislit mask in excellent observing conditions. 10 targets were initially selected from membership analysis based on full astrometric data from {\it Gaia} (this selection is further explained in Section \ref{subsec:gaia}). The remaining 49 targets were selected from stars that were consistent with the observed stellar population of UMa3/U1 in $r, i$ photometry which filled the slit mask around UMa3/U1.

We used the 1200G grating that covers a wavelength range of $6400-9100$\,\AA\ with the OG550 blocking filterwith the aim of measuring the Calcium\,II infrared triplet (CaT) absorption feature. Observations consisted of 3\,$\times$\,20\,minute exposures, or 3600s of total exposure time. Data were reduced to one dimensional wavelength-calibrated spectra using the open-source python-based data reduction code {\tt PypeIt} \citep{Prochaska20}.  {\tt PypeIt} reduces the eight individual DEIMOS detectors as four mosaic images, where each red/blue pair of detectors is reduced independently. When reducing data for this paper, {\tt PypeIt}’s default heliocentric and flexure corrections are turned off and a linear flexure term is determined as part of the 1D data reductions below.

Stellar radial velocities and Calcium Triplet equivalent widths (EWs) were measured using a preliminary version of the \textsc{DMOST} package (M. Geha et al., in prep). In brief, \textsc{DMOST} forward models the 1D stellar spectrum for each star from a given exposure with both a stellar template from the \textsc{PHOENIX} library and a telluric absorption spectrum from \textsc{TelFit} \citep{Gullikson14}. The velocity is determined for each science exposure through an MCMC procedure constraining both the radial velocity of the target star as well as a wavelength shift of the telluric spectrum needed to correct for slit miscentering \citep[see, e.g.][]{Sohn07}. The final radial velocity for each star is derived through an inverse-variance weighted average of the velocity measurements from each exposure. The systematic error reported by the pipeline, derived from the reproducibility of velocity measurements across masks and validated against spectroscopic surveys, is $\sim$\,1\,\kms\ (see M. Geha et al., in prep). Lastly, \textsc{DMOST} measures the equivalent width from the CaT by fitting a Gaussian-plus-Lorentzian model to the coadded spectrum (for stars at S/N\,$>$\,15) or a Gaussian model (for stars below S/N\,$<$\,15). We assume a 0.2 Angstrom systematic error on the total equivalent width determined from independent repeat measurements. Thanks to excellent observing conditions, we measured velocities and combined CaT equivalent widths (EW) for 31 of 59 targets and achieved spectra with a median signal-to-noise (S/N) per pixel ranging from 15 to 74 for the 10 suspected members.

%The final stellar membership determination will involve both {\it Gaia} proper motions (Section \ref{subsec:gaia}) as well as cuts in velocity and metallicity (Section \ref{subsec:spec}). Measured and derived quantities for member stars are presented in Table \ref{tab:stars}.

\section{Results}\label{sec:results}

\subsection{Distance and Stellar Population}\label{subsec:dist}

The matched-filter detection algorithm only searches for stellar populations with an age of 12\,Gyr and metallicity of \met\,= $-2$, and results of the search provide an initial estimate of the distance to UMa3/U1. 
We further refine these properties through visual inspection in an iterative process, where we make small systematic tweaks to the distance, metallicity, and age of the stellar population until a satisfactory model is settled upon. We also point out that UMa3/U1 has an extraordinarily low stellar mass (which will be explored further in Section \ref{subsec:mass}) so this by-eye analysis is based on only a handful of stars spanning $\sim6.5$\,mag on a color-magnitude diagram (CMD).

\begin{figure}
    \centering
    \includegraphics[width=\linewidth]{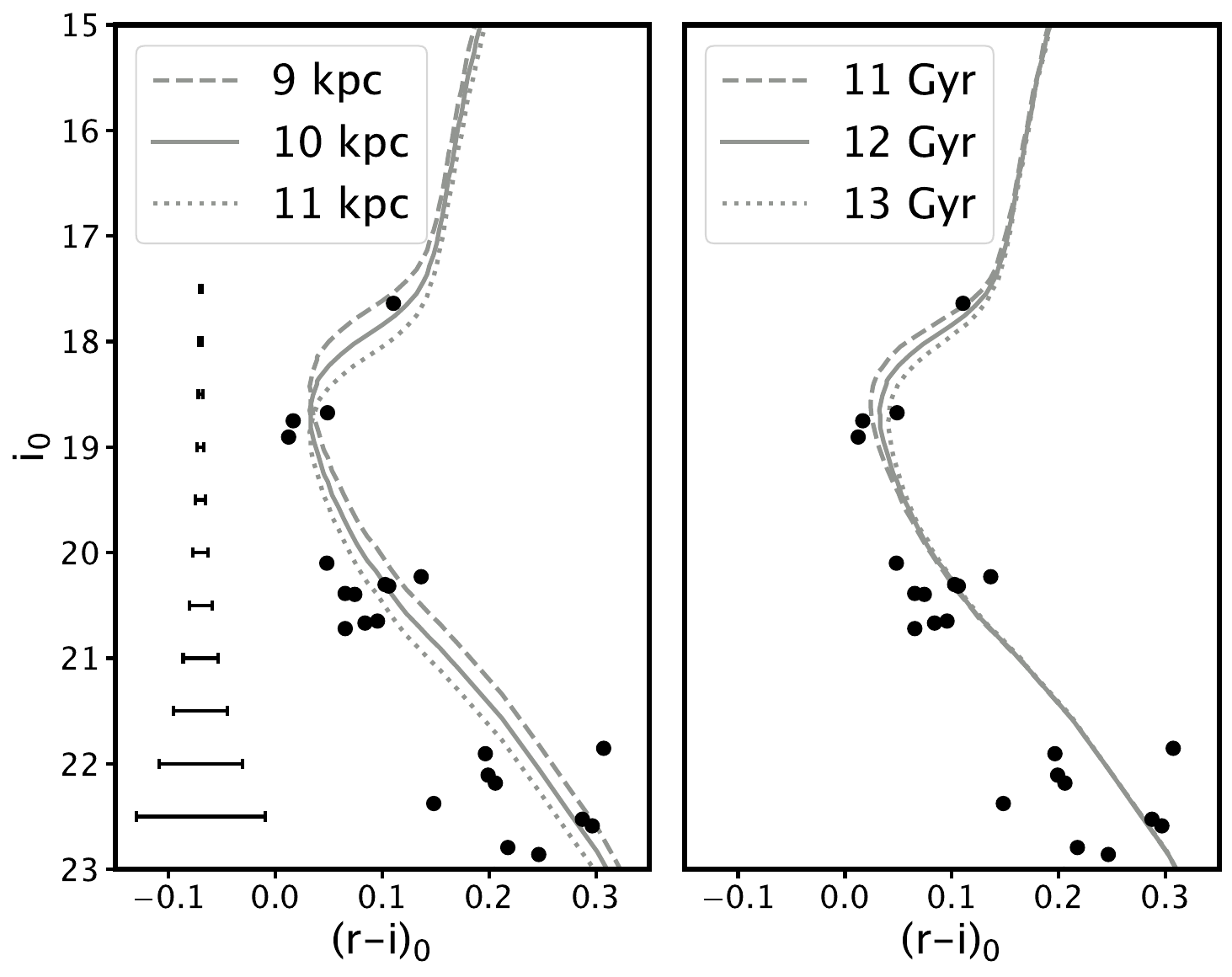}
    \caption{
    \plot{Left} CMD of all matched-filtered stars within 4\,$\times$\,\rh\ with a set 12\,Gyr, [Fe/H] = $-2.2$ isochrones plotted over top, shifted to distances of 9, 10, \& 11\,kpc to show the 10\% uncertainty assigned to the distance estimate. The median photometric uncertainties as a function of $i$ magnitude are shown in black on the left side of the CMD.
    \plot{Right} CMD same stars as on the left, with a set of [Fe/H] $= -2.2$ isochrones shifted to a distance of 10\,kpc where the age of the stellar population is varied between 11, 12, \& 13\,Gyr, showing that all such isochrones approximate the data reasonably well. 
    }
    \label{fig:agedist}
\end{figure}

We initially set the distance to be 10\,kpc and explored a grid of PARSEC isochrones, spanning $5 - 13$\,Gyr in age ($\tau$) in steps of 1\,Gyr and $-2.2$ to $-1.0$ in metallicity ([Fe/H]) in steps of 0.1\,dex by eye. \met\ = $-2.2$ appears to be a good fit to the data. $\tau \geq 11$\,Gyr is a good fit to the few stars at the Main-Sequence Turn Off (MSTO). Changing the age by a $\sim 1$\,Gyr results in minor changes in the shape of the isochrone, while changes in metallicity are negligible, so we adopt $\tau = 12$\,Gyr and [Fe/H] $= -2.2$ for all calculations and derivations going forward. We note, however, that [Fe/H] $= -2.2$ is the most metal-poor isochrone available in our set, so this isochrone may only be an upper limit on the true systemic metallicity.

For the distance estimate, we prioritize fitting the MSTO to the four brightest stars due to their small photometric uncertainties, though the fainter stars appear to be slightly bluer on average than our best-fit isochrone, the most metal-poor in the set. While this could be explained by the large photometric uncertainties at this magnitude, we overlaid slightly more metal-poor MIST isochrones \citep{Choi16}. We did not find a significantly better fit, and note that variance between isochrones from different databases is large enough that attempting to constrain the distance with greater accuracy may not be appropriate. We therefore adopt a 10\% uncertainty on the distance estimate, giving 10\,$\pm$\,1\,kpc. Adjustments of 1\,kpc were incrementally made to the distance of the isochrone, and this appears to be appropriate. 
Figure \ref{fig:agedist} shows CMDs with all matched-filter-selected stars brighter than $i_0 \sim 23$\,mag, where PARSEC isochrones are plotted over top with small variations in distance and age.

Unfortunately, we are unable to use either Tip of the Red Giant Branch (TRGB) or Horizontal Branch (HB) stars to further constrain the distance to UMa3/U1 as this system is so sparsely populated that no stars currently exist in these more highly evolved, shorter duration stages of the stellar lifecycle. Given the relative closeness of the system, along with the $r$ saturation limits of $r \sim 17.5$\,mag, we also searched through the Third Data Release (DR3) from {\it Gaia} \citep{Gaia16, Gaia21}, but a lack of bright stars in the direction of UMa3/U1 confirms that no TRGB or HB stars are present. Additionally, we searched the PS1 RR Lyrae \citep{Sesar17} and {\it Gaia} variability \citep{Gaia22} catalogs, but could not identify any RR Lyrae stars within a few arcminutes of UMa3/U1. We therefore cannot use the properties of these variable stars to further constrain the heliocentric distance.

\subsection{Structural Parameters}\label{subsec:structural}

\begin{deluxetable}{lc}
\tabletypesize{\footnotesize}
\tablecaption{Flat priors for each parameter in the MCMC analysis \label{tab:priors}}

\tablehead{
    Parameter & Prior
}

\startdata
$x_0$ & $-6$\arcmin $\leq$ $\Delta x_0$ $\leq$ $+6$\arcmin \\
$y_0$ & $-6$\arcmin $\leq$ $\Delta y_0$ $\leq$ $+6$\arcmin \\
\rh & 0 $<$ \rh\ $<$ $9$\arcmin \\
$\epsilon$ & 0 $\leq$ $\epsilon$ $<$ 1 \\
$\theta$ & $-$90\textdegree\ $\leq$ $\theta$ $<$ $+$90\textdegree \\
N$^*$ & 0 $\leq$ $N^*$ $\leq$ 100 \\
\enddata

\end{deluxetable}

We follow a procedure, which is based on the methodology laid out in \citet{Martin08, Martin16-pds}, to estimate the structural parameters of UMa3/U1 assuming the distribution of member stars is well described by an elliptical, exponential radial surface density profile and a constant field contamination. The profile, $\rho_{\text{dwarf}}(r)$, is parameterized by the centroid of the profile ($x_0, y_0$), the ellipticity $\epsilon$ (defined as $\epsilon = 1 - b/a$ where $b/a$ is the minor-to-major-axis ratio of the model), the position angle of the major axis $\theta$, (defined East of North), the half-light radius (which is the length of the semi-major axis \rh), and the number of stars $N^*$ in the system. The model is written as 

\begin{equation}
	\rho_{\text{dwarf}}(r) = \frac{1.68^2}{2\pi\rhm^2(1-\epsilon)}N^*\exp\bigg(\frac{-1.68r}{\rhm}\bigg)
\end{equation}

\noindent where $r$, the elliptical radius, is related to the projected sky coordinates ($x$, $y$) by

\begin{eqnarray}
    \nonumber r = \Bigg\{ \bigg[\frac{1}{1-\epsilon} \Big((x-x_0)\cos\theta - (y-y_0)\sin\theta\Big) \bigg]^2 \\ + \bigg[ \Big((x-x_0)\sin\theta - (y-y_0)\cos\theta\Big)^2 \bigg]^2\Bigg\}^{\frac{1}{2}}.
\end{eqnarray}

\noindent We assume that the background stellar density is uniform, which is reasonable on the scale of arcminutes up to a degree or so. The background, $\Sigma_{\text{b}}$, is calculated as follows:

\begin{equation}
    \Sigma_{\text{b}} = \frac{n - N^*}{A},
\end{equation}

\noindent where $n$ is the total number of stars in the field of view and $A$ is the total area, normalizing the background density with respect to the selected region. We combine the elliptical, exponential surface density model with the uniform background term to construct our posterior distribution function,

\begin{equation}
    \rho_{\text{model}}(r) = \rho_{\text{dwarf}}(r) + \Sigma_{\text{b}},
\end{equation}

\noindent which we sample with the the afine-invariant Markov Chain Monte Carlo sampler \texttt{emcee} \citep{Foreman13} to estimate the most likely structural parameters to describe UMa3/U1. We apply flat priors to all parameters, the bounds of which are given in Table \ref{tab:priors}.

Prior to invoking this method, we select only those stars which are selected by the matched-filter, this time using the 12\,Gyr, [Fe/H] = $-$2.2 isochrone which was determined in Section \ref{subsec:dist}. We apply an additional magnitude cut, retaining stars with $i \leq 23.5$\,mag to ensure incompleteness effects do not skew the parameter estimation.

The \texttt{emcee} routines are initialized with 64 walkers, each running for 10,000 iterations with the first 2,000 iterations thrown out to account for burn-in. The program shows good convergence and the median value for each parameter, along with the 16\% and 84\% percentiles taken as uncertainties, are provided in Table \ref{tab:props} along with all other measured and derived properties. UMa3/U1 is compact, with a physical half-light radius (\rh, semi-major axis of elliptical distribution) derived to be 0.9$^{+0.4}_{-0.3}$\arcmin, or 3\,$\pm$\,1\,pc in physical units.

\begin{deluxetable*}{clc}
\tabletypesize{\footnotesize}
\tablecaption{Measured and derived properties for Ursa Major 3/UNIONS 1 \label{tab:props}}

\tablehead{
    Property & Description & Value
}

\startdata
$\alpha_{J2000}$ & Right Ascension & 11$h$ 38$m$ 49.8$s$ \\
$\delta_{J2000}$ & Declination & +31\textdegree\ 4\arcmin\ 42\arcsec \\
$r_{\text{h,ang}}$ & Angular Half-Light Radius & 0.9$^{+0.4}_{-0.3}$\arcmin \\
$r_{\text{h,phys}}$ & Physical Half-Light Radius & 3\,$\pm$\,1\,pc \\
$\epsilon$ & Ellipticity & 0.5$^{+0.2}_{-0.3}$ \\
$\theta$ & Position Angle & 169$^{+18}_{-12}$\,deg \\
N$^*$ & Number of Stars (down to $i = 23.5$\,mag) & 21$^{+6}_{-5}$ \\
$D_{\odot}$ & Heliocentric Distance & 10\,$\pm$\,1\,kpc\\
$(m-M)_0$ & Distance Modulus & 15.0\,$\pm$\,0.2\,mag\\
$\tau$ & Age (Isochrone) & $12$\,Gyr\tablenotemark{a}\\
\met & Metallicity (Isochrone) & $-2.2$\,dex\tablenotemark{b} \\
M$_{\text{tot}}$ & Total Stellar Mass & 16$^{+6}_{-5}$\,\Msun \\
$M_V$ & Absolute $V$-band Magnitude & +2.2$^{+0.4}_{-0.3}$\,mag \\
N$_{\text{tot}}$ & Total Number of Stars & 57$^{+21}_{-19}$ \\
$\mu_{\text{eff}}$ & Effective Surface Brightness & 27\,$\pm$\,1\,mag\,arcsec$^{-2}$ \\
\midrule
$\mu_{\alpha}$ cos $\delta$ & Proper Motion in R.A. & $-$0.75\,$\pm$\,0.09\,(stat)\,$\pm$\,0.033\,(sys)\,mas\,year$^{-1}$\tablenotemark{c} \\
$\mu_{\delta}$ & Proper Motion in Dec. & 1.15\,$\pm$\,0.14\,(stat)\,$\pm$\,0.033\,(sys)\,mas\,year$^{-1}$\tablenotemark{c}\\
$\langle v_{\odot}\rangle$ & Mean Heliocentric Velocity & 88.6\,$\pm$\,1.3\,\kms \\
$\sigma_{v}$ & Intrinsic Velocity Dispersion & 3.7$^{+1.4}_{-1.0}$\,\kms\tablenotemark{d} \\
M/L$_{1/2}$ & Dynamical Mass-to-Light Ratio ($< r_{\text{h}})$ & 6500$^{+9100}_{-4300}$\,\Msun/\Lsun\tablenotemark{d} \\
$r_{\text{peri}}$ & Pericenter & 12.9$^{+0.7}_{-0.7}$\,kpc \\ 
$r_{\text{apo}}$ & Apocenter & 26$^{+2}_{-3}$\,kpc \\
$z_{\text{max}}$ & Maximum height above MW Disk & 17$^{+2}_{-2}$\,kpc \\
$\epsilon_{\text{orb}}$ & Orbital Eccentricity & 0.34$^{+0.02}_{-0.03}$ \\
$\Delta t_{\text{peri}}$ & Time Between Pericenters & 373$^{+32}_{-34}$\,Myr \\
$t_{\text{since}}$ & Time Since Last Pericenter & 31$^{+2}_{-2}$\,Myr \\
\enddata

\tablenotetext{a}{All isochrones with $\tau > 11$\,Gyr fit the data well, but we adopt $\tau = 12$\,Gyr for computations.}
\tablenotetext{b}{[Fe/H] = $-2.2$ is the most metal-poor isochrone available in our set.}
\tablenotetext{c}{Systematic uncertainties were investigated by \citet{Lindegren21}.}
\tablenotetext{d}{The removal of a single star (\#2 in Table \ref{tab:stars}) drops the velocity dispersion to $1.9^{+1.4}_{-1.1}$\,\kms, decreasing the inferred M/L$_{1/2}$ to 1900$^{+4400}_{-1600}$\,\Msun/\Lsun.}

\end{deluxetable*}

\begin{figure*}
    \centering
    \includegraphics[width=0.8\linewidth]{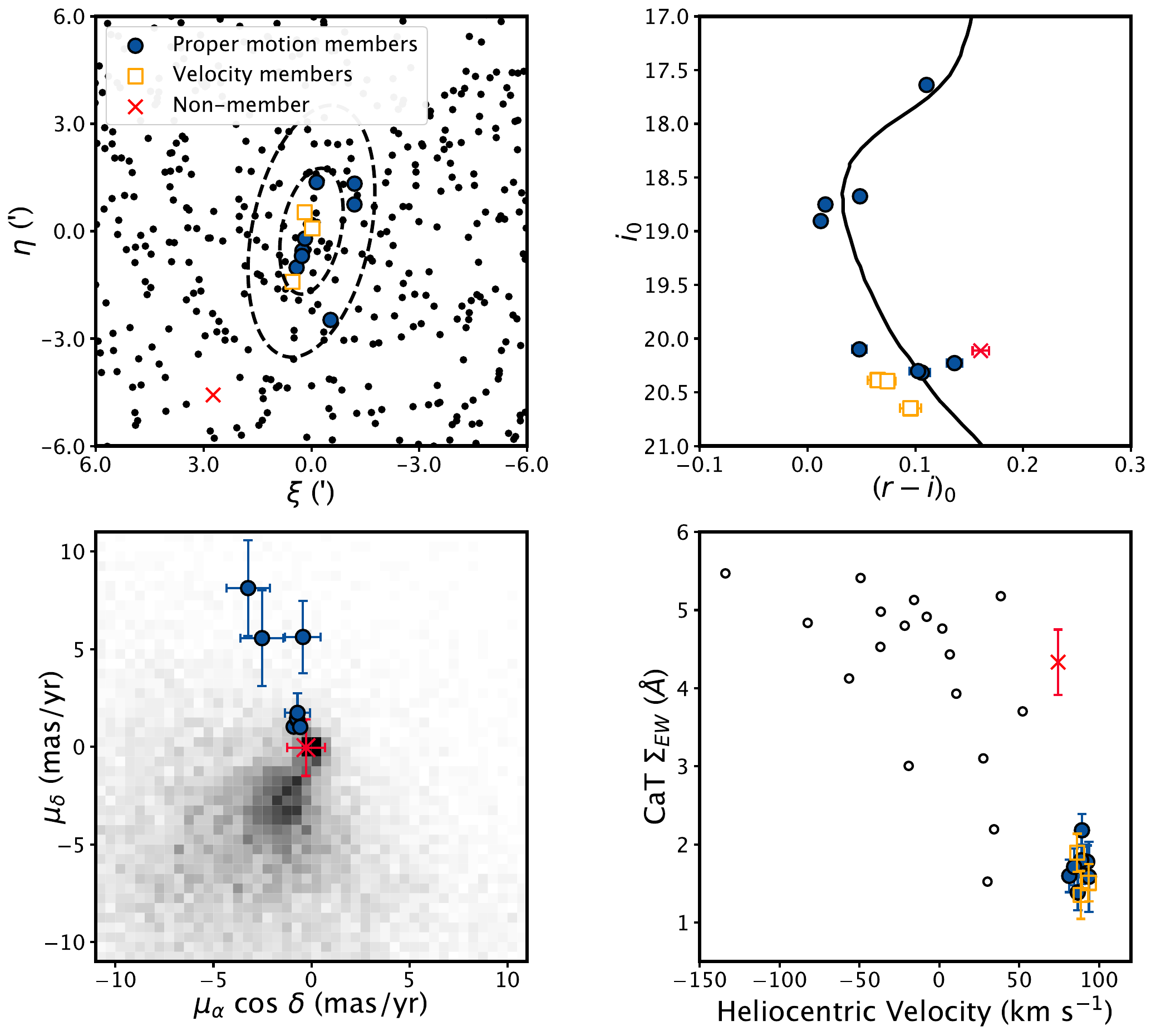}
    \caption{
    \plot{Top Left} 12\arcmin\,$\times$ 12\,\arcmin\ region around the UMa3/U1. Blue markers are high likelihood members ($P_{\text{sat}} \geq 0.75$) as determined with membership analysis code using {\it Gaia} proper motions. Orange square markers do not have full astrometric measurements, but are members based on velocity and CaT EW. Red `X' has marginal membership likelihood from {\it Gaia} and is likely not a member based on Keck/DEIMOS measurements. Black sources are all those selected by the matched filter in UNIONS (i.e. same as the blue sources in Figure \ref{fig:cmd-detect}).
    \plot{Top Right} Suspected member stars observed with Keck/DEIMOS plotted on a UNIONS $r,i$ extinction-corrected CMD, with a 12\,Gyr, [Fe/H] = $-2.2$ isochrone overlaid, shifted to a distance of 10\,kpc. Color uncertainties are shown, though they are only just visible for the sources around $i_0 \sim 20.5$\,mag.
    \plot{Bottom Left} Proper motion measurements from {\it Gaia} where coloring is the same as on the CMD. The underlying grey density plot is Milky Way foreground proper motion distribution, measured empirically from stars within 2\,deg of UMa3/U1. Note that 5 of the likely members are all tightly clustered around the systemic proper motion at ($\mu_{\alpha}$cos$\delta$, $\mu_{\delta}$) = ($-$0.75, 1.15)\,mas\,year$^{-1}$.
    \plot{Bottom Right} Heliocentric velocity plotted against CaT $\Sigma_{\text{EW}}$ where coloring is the same as on CMD, and small black circles are other non-member stars observed with Keck/DEIMOS. See Figure \ref{fig:veldist} for a closer look at the velocity distribution near the mean systemic velocity.
    }
    \label{fig:gaiacmd}
\end{figure*}

\begin{figure}
    \centering
    \includegraphics[width=\linewidth]{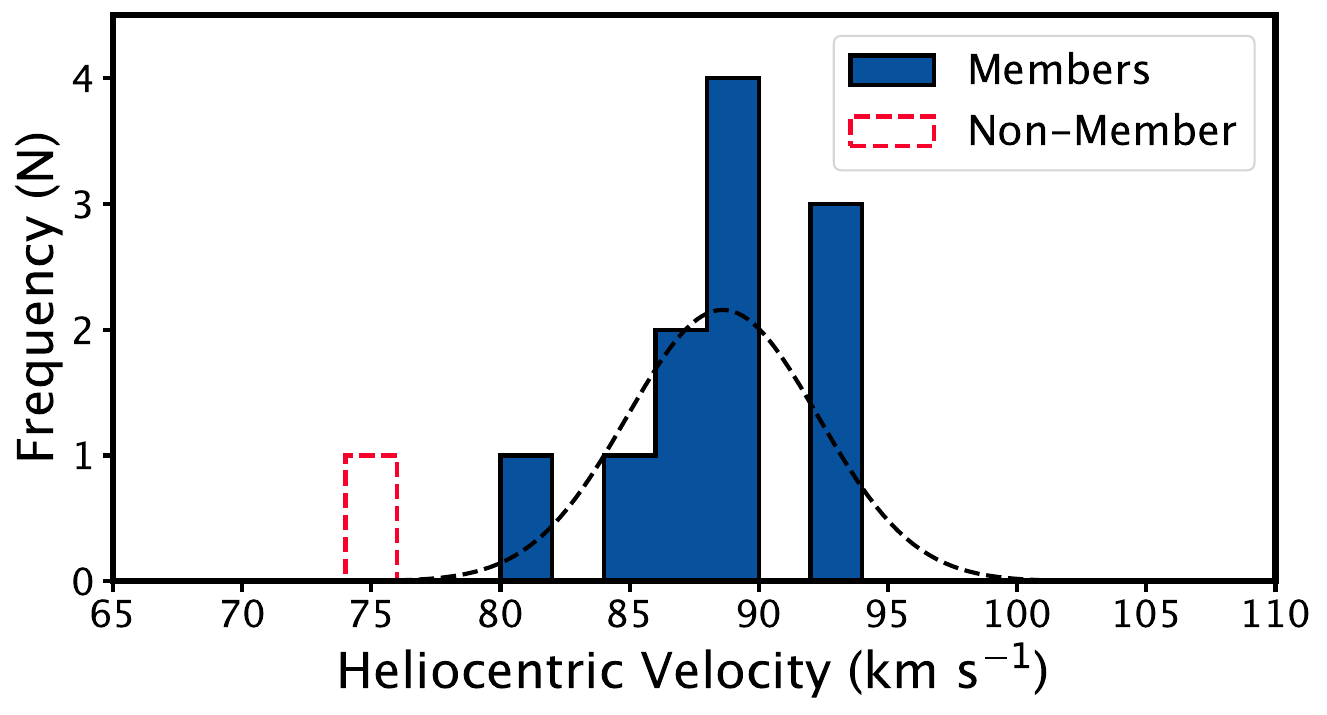}
    \caption{
    Velocity distribution of candidate member stars (dark blue) and a star with marginal membership probability (dashed red, $P_{\text{sat}} = 0.02$). 
    The black dashed line is the velocity probability distribution function with a dispersion of 3.7\,\kms\ as derived in Section \ref{subsec:velocity}.
    }
    \label{fig:veldist}
\end{figure}

\subsection{Proper Motion}\label{subsec:gaia}

Despite its extreme dearth of giant stars, UMa3/U1 lies close enough that several stars are brighter than the approximate limiting magnitude ($G \sim 21$\,mag) of \textit{Gaia} DR3. Using stars with full astrometric data \citep{Lindegren21}, we estimate the systemic proper motion of UMa3/U1 and assign likelihoods for individual stars to be members of this faint stellar system.

We follow the methodology of \citet{Jensen23}, which builds upon \citet{McVenn20a, McVenn20b}, and we refer the reader to those papers for more details regarding the algorithm. Briefly, the method uses spatial, photometric, and astrometric information about each star, in conjunction with the structural parameters derived for some stellar system, to compute the likelihood that a given star is a member of a putative stellar system rather than a member of the Milky Way foreground stellar population. The Milky Way foreground prior distributions for each parameter are calculated empirically using a subset of stars detected by {\it Gaia} in a 2\,deg circle about the location of the system, where the {\it Gaia} photometry was extinction-corrected following \citet{Gaia18}.

In this application to UMa3/U1, we use the structural parameters derived from UNIONS data in Section \ref{subsec:structural} and the stellar population estimates of $\tau = 12$\,Gyr and [Fe/H] $= -2.2$. 
The systemic proper motion of UMa3/U1 is estimated by the membership algorithm to be $(\mu_\alpha$cos$\delta, \mu_\delta) = (-0.75$\,$\pm$\,0.09\,(stat)\,$\pm$\,0.033\,(sys), $1.15$\,$\pm$\,0.14\,(stat)\,$\pm$\,0.033\,(sys)) mas\,year$^{-1}$.
This process also identifies 7 stars with $P_{\text{sat}} > 0.99$ (where $P_{\text{sat}}$ is membership likelihood) and an additional star with $P_{\text{sat}} = 0.75$. \citet{Jensen23} have investigated the workings of this code, finding that $P_{\text{sat}} \sim 0.2$ actually corresponds to $\sim$\,50\% probability of being a member (based on known members identified via radial velocities), so we find 8 high-likelihood member stars in total based on {\it Gaia} measurements. 2 additional stars with marginal membership likelihoods ($0.01 < P_{\text{sat}} < 0.10$) near to the centroid of UMa3/U1 were also targeted in our spectroscopic follow-up for further investigation.

\subsection{Membership}\label{subsec:mems}

As discussed in Section \ref{subsec:keck}, 31 of 59 target spectra were successfully measured following Geha et al. (in prep.), producing heliocentric radial velocities (\vh) and combined CaT EWs (sum of all three CaT absorption features, $\Sigma_{\text{EW}}$). Traditionally, methods of inferring metallicity ([Fe/H]) from CaT EWs have been calibrated using red giant branch stars \citep{Starkeburg10, Carrera13}. Due to its extraordinarily low stellar mass, U1 only has a single post-MS star visible on its CMD, with this one star lying at the inflection point of the best fitting isochrone where the sub-giant branch (SGB) transitions to the red giant branch. We estimate this star to have log($g$) = 3.51 using the best-fit isochrone, so while it has evolved off the MS, it is outside the range of log($g$) for which [Fe/H] estimators have been calibrated. The calibration from \citet{Starkeburg10} considers model spectra computed for RGBs with $0.5 \leq $\ log($g$) $\leq 2.5$ while \citet{Carrera13} calibrated the [Fe/H] $-$ CaT $\Sigma_{\text{EW}}$ relationship using RGBs that lay in the range $0.7 \leq $\ log($g$) $\leq 3.0$, where log($g$) was derived from high resolution spectroscopy. However, we still are able to use the measured $\Sigma_{\text{EW}}$ to help with membership identification.

Using a combination of \vh, $\Sigma_{\text{EW}}$, and membership likelihoods assigned from the algorithm described in Section \ref{subsec:gaia}, we isolate member stars.
In Figure \ref{fig:gaiacmd}, we plot all potential member stars on-sky and on a CMD of extinction-corrected $r, i$ photometry. The four brightest members are all in good agreement with the best-fit isochrone. In the range $20 \leq G_0 \leq 21$\,mag, there are several stars that are all roughly consistent with the isochrone, though there is some scatter. 
The eight sources marked with blue circles are those with high membership likelihoods ($P_{\text{sat}} \geq 0.75$) and the orange squares are three additional stars that are consistent with {\it Gaia}-identified members both spatially and in \vh, but do not have full astrometric data, and could therefore not be identified by the membership algorithm. The red `X' demarcates one of the marginal members ($P_{\text{sat}} = 0.02$) noted in Section \ref{subsec:gaia} and its formal membership will be discussed below (Section \ref{subsubsec:marg}).
All 8 member stars are seen to cluster in proper motion space (bottom-left panel of Figure \ref{fig:gaiacmd}), and we note that 5 of these members are tightly clustered at the systemic proper motion (see Table \ref{tab:props}). The other 3 other members have large uncertainties, as they are the faintest of the members with measured proper motions, but are still consistent with the systemic proper motion (within $2-3\sigma$). Additionally, the bottom-right panel of Figure \ref{fig:gaiacmd} shows the $\Sigma_{\text{EW}}$ $-$ \vh\ plane of all stars observed with Keck/DEIMOS. The suspected members cluster at $v_{\odot} \sim 90$\,km\,s$^{-1}$ and $\Sigma_{\text{EW}}$ $\sim$\,1.5\,\AA, neatly separated from all other measured stars. Photometric and dynamical properties are listed for all likely members and the two confirmed non-members in Table \ref{tab:stars}.

\begin{deluxetable*}{ccccccccccc}
\tabletypesize{\footnotesize}
\tablecaption{Candidate Member Stars Targeted by Keck/DEIMOS \label{tab:stars}}
\tablehead{
    \colhead{Object} & 
    \colhead{Gaia Source ID} & 
    \colhead{RA (J2000)} & 
    \colhead{DEC (J2000)} & 
    \colhead{$r_0$} & 
    \colhead{$i_0$} & 
    \colhead{$v_{\odot}$} & 
    \colhead{CaT $\Sigma_{\text{EW}}$} & 
    \colhead{S/N} & 
    \colhead{$P_{\text{sat}}$} & 
    \colhead{Member?} \\
    \nocolhead{} & 
    \nocolhead{} & 
    \colhead{(deg)} & 
    \colhead{(deg)} & 
    \colhead{(mag)} & 
    \colhead{(mag)} & 
    \colhead{(\kms)} & 
    \colhead{(\AA)} & 
    \colhead{(pixel$^{-1}$)} & 
    \nocolhead{} &
}

\startdata
1 & 4024083571202406912 & 174.69706 & 31.03707 & 17.75 & 17.64 & 89.3\,$\pm$\,1.3 & 2.2\,$\pm$\,0.2 & 74 & 0.99 & Y \\
2 & 4024177442007708672 & 174.71544 & 31.06145 & 18.72 & 18.68 & 81.4\,$\pm$\,1.5 & 1.6\,$\pm$\,0.2 & 43 & 1.0 & Y \\
3 & 4024177648166139904 & 174.71228 & 31.06923 & 18.77 & 18.75 &  89.9\,$\pm$\,1.5 & 1.8\,$\pm$\,0.2 & 41 & 1.0 & Y \\
4 & 4024178472800038016 & 174.68413 & 31.09085 & 18.92 & 18.91 &  92.6\,$\pm$\,1.6 & 1.8\,$\pm$\,0.2 & 37 & 0.99 & Y \\
5 & 4024179297433597056 & 174.70462 & 31.15346 & 19.73 & 19.83 &  n/a\tablenotemark{a} & n/a\tablenotemark{a} & 20 & 0.10 & N \\
6 & 4024177648166141184 & 174.71091 & 31.07493 & 20.15 & 20.10 &  88.9\,$\pm$\,2.0 & 1.8\,$\pm$\,0.3 & 19 & 0.99 & Y \\
7 & 4024178472800038144 & 174.68406 & 31.10051 & 20.36 & 20.23 &  86.7\,$\pm$\,2.5 & 1.4\,$\pm$\,0.2 & 16 & 0.99 & Y \\
8 & 4024036262137953536 & 174.76056 & 31.00214 & 20.27 & 20.11 &  74.4\,$\pm$\,1.9 & 4.3\,$\pm$\,0.4 & 16 & 0.02 & N \\
9 & 4024177442007709440 & 174.71257 & 31.06684 & 20.40 & 20.30 &  93.7\,$\pm$\,2.5 & 1.6\,$\pm$\,0.4 & 16 & 0.75 & Y \\
10 & 4024177751245359744 & 174.70460 & 31.10120 & 20.42 & 20.32 &  84.4\,$\pm$\,2.6 & 1.7\,$\pm$\,0.2 & 15 & 0.99 & Y \\
11 & 4024177682525881344 & 174.71106 & 31.08719 & 20.45 & 20.39 &  93.6\,$\pm$\,2.5 & 1.5\,$\pm$\,0.2 & 15 & n/a\tablenotemark{b} & Y \\
12 & 4024177442007706752 & 174.71763 & 31.05481 & 20.47 & 20.40 &  86.3\,$\pm$\,2.8 & 1.9\,$\pm$\,0.2 & 15 & n/a\tablenotemark{b} & Y \\
13 & 4024177648166141312 & 174.70698 & 31.07972 & 20.74 & 20.65 &  88.6\,$\pm$\,3.3 & 1.4\,$\pm$\,0.3 & 13 & n/a\tablenotemark{b} & Y
\enddata

\tablenotetext{a}{Upon further inspection of the extracted 1D spectra, this source may be a distant quasar, as there are no CaT absorption features and a broad emission-like bump at $\sim$\,7200\,\AA. The best-fit model spectrum is not informative, and therefore the measured velocity is excluded.}

\tablenotetext{b}{These three stars do not have full astrometric measurements in {\it Gaia} DR3 and therefore $P_{\text{sat}}$ could not be computed, but are nonetheless suspected to be members based on their velocities and CaT equivalent widths.}

\end{deluxetable*}

\subsubsection{Marginal Members}\label{subsubsec:marg}

Here, we offer evidence to suggest that the two marginal members identified in Section \ref{subsec:gaia} ($0.01 < P_{\text{sat}} < 0.10$) are in truth not members of UMa3/U1. First, star \#5 in Table \ref{tab:stars} had very unusual measurements in both velocity and $\Sigma_{\text{EW}}$, so we examined its spectrum and found it to lack any CaT absorption features while featuring a broad emission-like bump around 7200\,\AA. Paired with a proper motion of close to zero, we suspect that this source may be a background quasar and consequently exclude it from our analysis.

Star \#8 in Table \ref{tab:stars} is featured in both Figures \ref{fig:gaiacmd} \& \ref{fig:veldist} as a red `X' where measurements show evidence that this is not a member. While this star does lie close to the systemic proper motion of UMa3/U1, it also has a proper motion near zero. In Figure \ref{fig:gaiacmd}, the grey contours show the empirical distribution of all stars measured by {\it Gaia} within 2\,deg of UMa3/U1 and there is an overdensity near the origin, giving this star a high likelihood of being part of the proper motion background. Additionally, this star is more than 7\,$\times$\,\rh\ from the centroid of the stellar overdensity.
Figure \ref{fig:veldist} shows that this star is offset from the mean velocity by 3.8$\sigma$, where the dispersion of the distribution is calculated in Section \ref{subsec:velocity}.
The final piece of evidence comes from the $\Sigma_{\text{EW}}$ relative to other likely member stars. Despite not being able to calculate [Fe/H], we can see on the {\it Gaia} CMD in Figure \ref{fig:gaiacmd} that there are 7 suspected members, as well as the marginal member being discussed here, in the range $20 \leq G_0 \leq 21$\,mag. Therefore, if all these stars are true members, they are all similar types of dwarf stars and thus should have similar values of log($g$), T$_{eff}$, and [Fe/H]. Additionally, the isochrone fitting implies that this is a metal-poor population ([Fe/H] $\sim -2.2$). The 7 suspected members are all clustered around $\sim$\,1.5\AA\ while star \#8 has $\Sigma_{\text{EW}}$ = 4.42\,$\pm$\,0.44\,\AA. If this star in question really is a main sequence dwarf star in UMa3/U1 (and therefore at the same distance of 10\,kpc), it must be far more metal-rich than than the other suspected member stars, which would be extremely anomalous. All told, we feel this accumulation of evidence implies that star \#8 is very unlikely to be a member of UMa3/U1, so we exclude it from the member list for the following analysis of the velocity distribution.

\begin{figure*}
    \centering
    \includegraphics[width=0.32\linewidth]{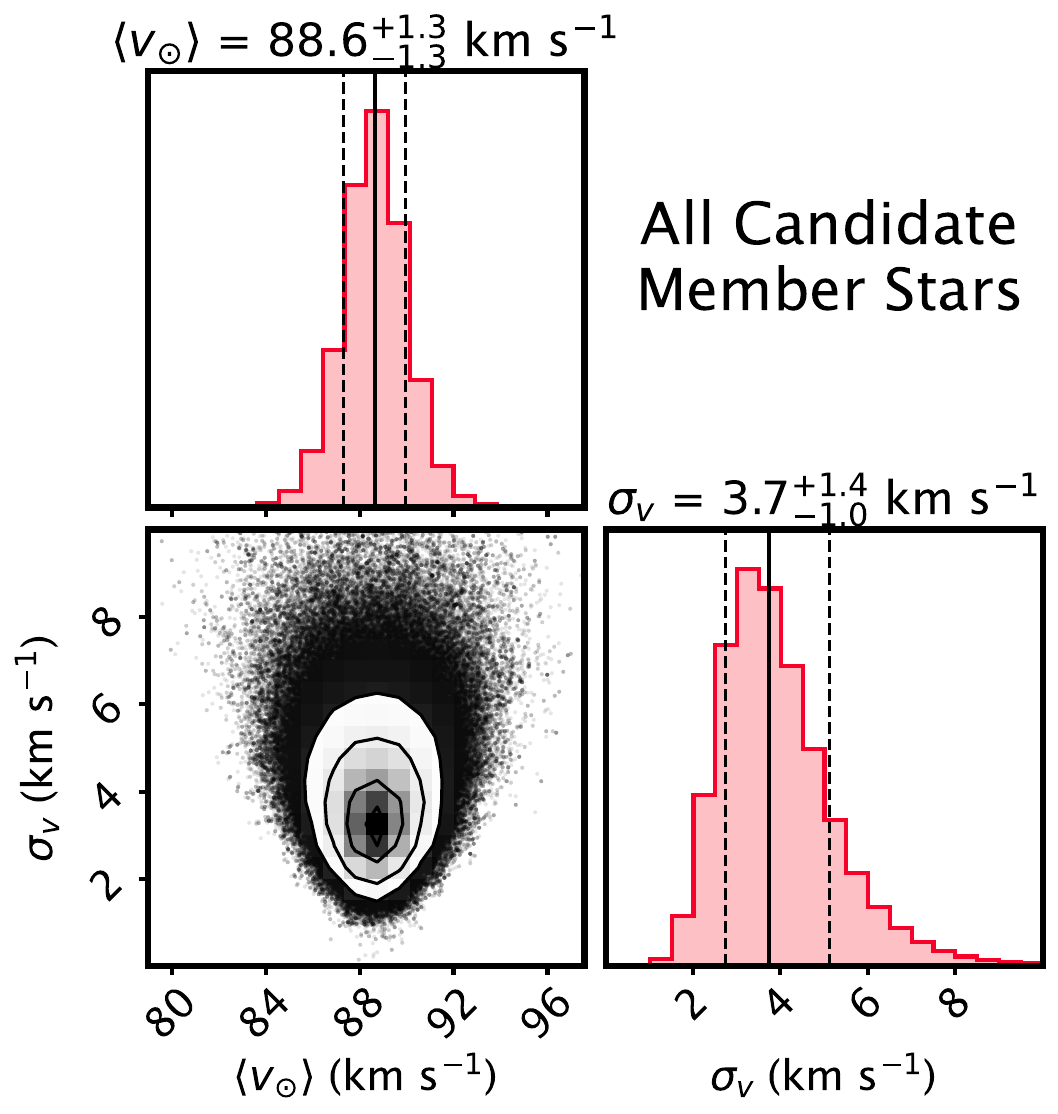}
    \includegraphics[width=0.32\linewidth]{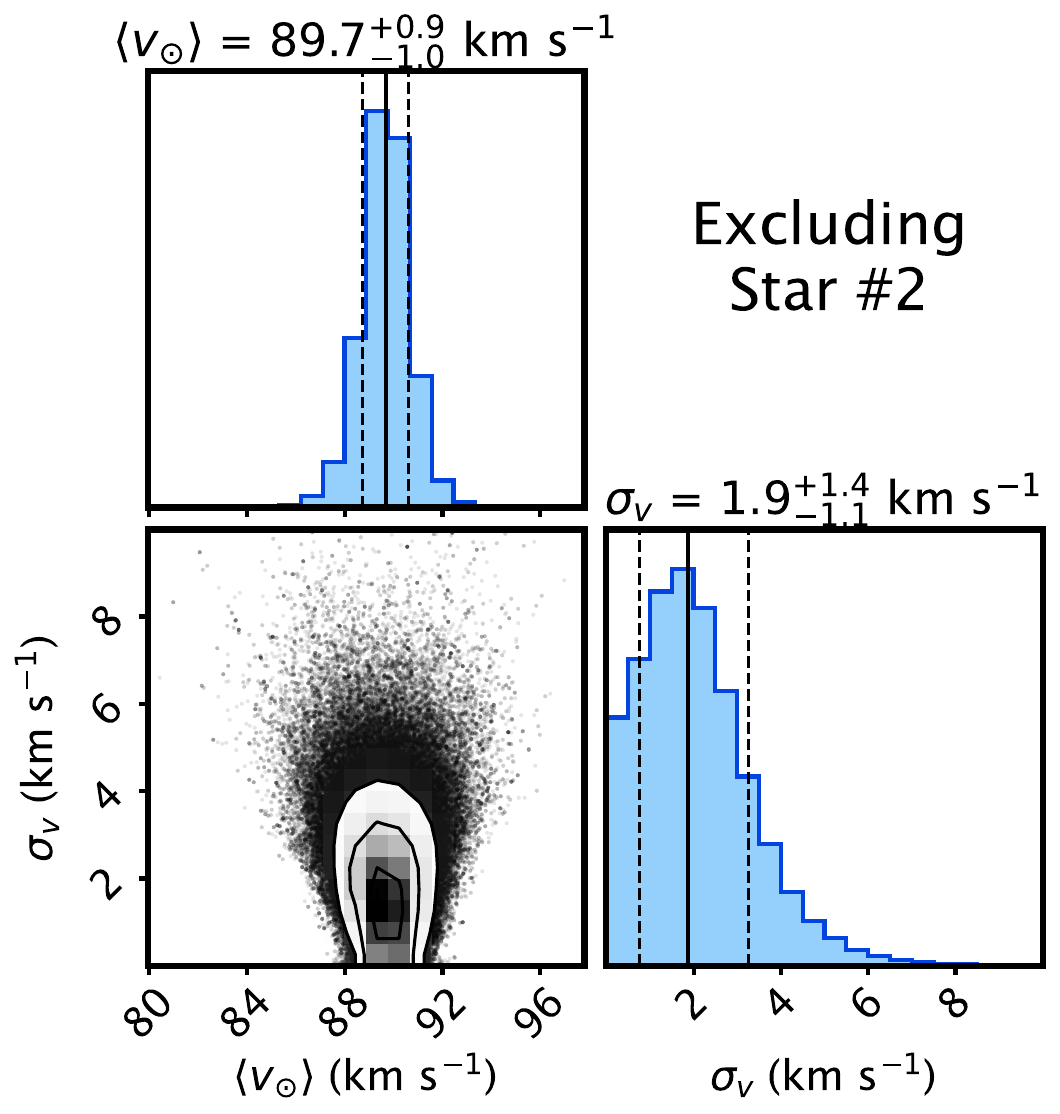}
    \includegraphics[width=0.32\linewidth]{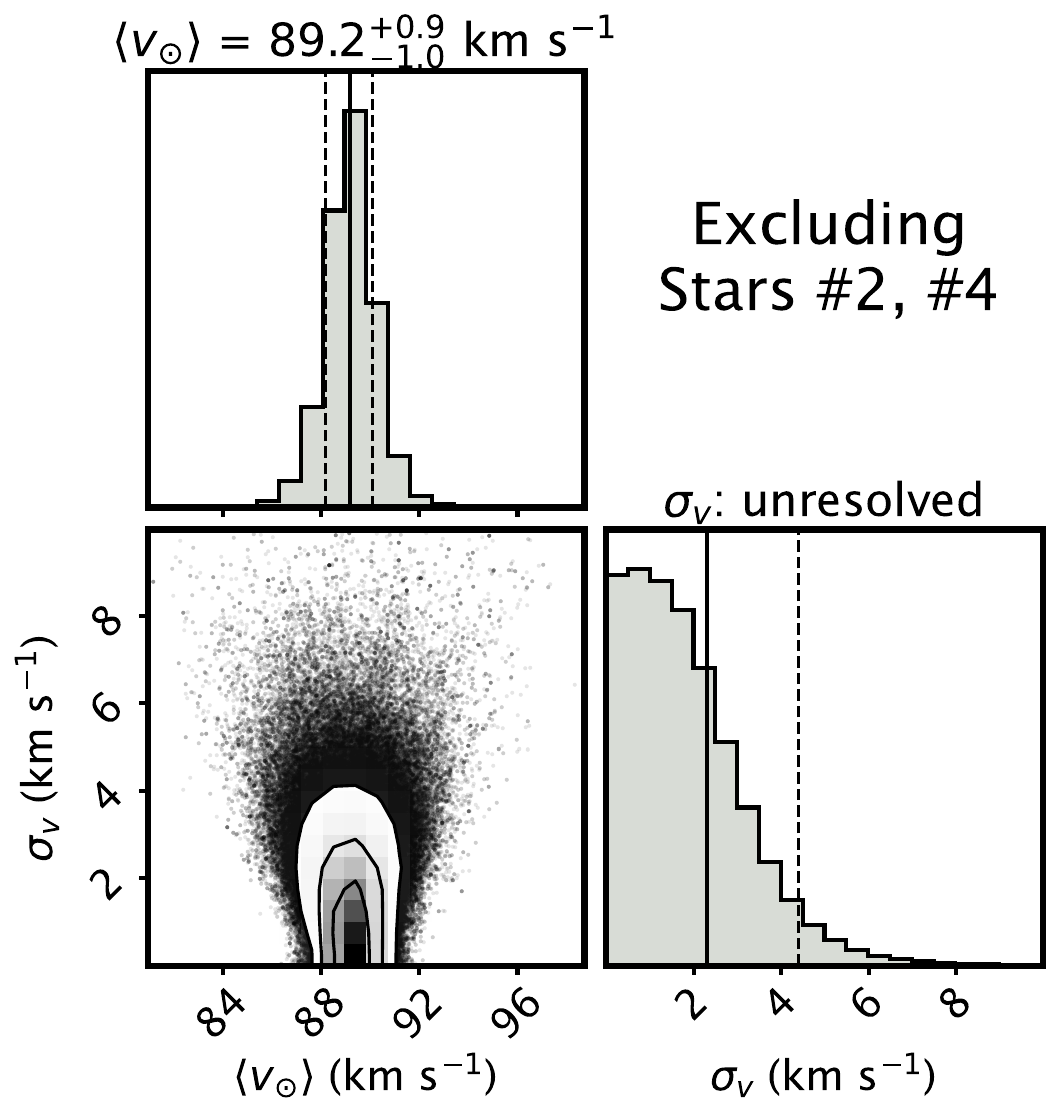}
    \caption{
    {\it Left}: 2D and marginalized posterior probability distributions for the mean heliocentric radial velocity and its intrinsic dispersion measured from 11 likely member stars. The median values of the heliocentric radial velocity and intrinsic dispersion are shown with uncertainties indicating the 16\% and 84\% percentiles.
    {\it Center}: Same as the left panel, except we have excluded star \#2 (in Table \ref{tab:stars}) and note that the intrinsic velocity dispersion drops to 1.9$^{+1.4}_{-1.1}$\,km\,s$^{-1}$. 
    {\it Right}: Same as the left panel, except we have excluded stars \#2 and \#4 (in Table \ref{tab:stars}). In this case, we no longer resolve an intrinsic velocity dispersion. Upon testing the systematic removal of all combinations of stars, the exclusion of stars \#2 and \#4 are the most impactful upon the velocity dispersion. We report 68\% and 95\% percentile upper limits on the velocity dispersion of 2.3 and 4.4\,\kms\ respectively.
    }
    \label{fig:mems-mcmc}
\end{figure*}

\subsection{Velocity Distribution}\label{subsec:velocity}

We follow \citet{Walker06} to measure the mean and intrinsic dispersion in heliocentric velocity. We construct the following log-likelihood function,

\begin{equation}
    \ln{L} = -\frac{1}{2}\sum^N_{i=1}\ln{(\sigma_i^2 + \sigma_v^2)} - \frac{1}{2}\sum^N_{i=1}\frac{(v_i - \langle v_{\odot}\rangle)^2}{\sigma_i^2 + \sigma_v^2} - \frac{N}{2}\ln(2\pi),
\end{equation}

\noindent where the $v_i$ and $\sigma_i$ are the measured velocity and uncertainty for each individual star. $\langle v_{\odot}\rangle$ is the mean heliocentric radial velocity of UMa3/U1, and $\sigma_v$ is the intrinsic velocity dispersion, which are the model parameters and quantities of physical interest. The log-likelihood function is maximized with respect $\langle v_{\odot}\rangle$ and $\sigma_v$ using \texttt{emcee}. The final estimates are $\langle v_{\odot}\rangle = 88.6$\,$\pm$\,$1.3$\,\kms\ and $\sigma_v = 3.7^{+1.4}_{-1.0}$\,\kms\ where uncertainties are the 16th and 84th percentiles of the distributions produced by the MCMC. This result is shown in the left-most panels of Figure \ref{fig:mems-mcmc}.

We investigate the robustness of this result, to understand how the selection of UMa3/U1 member stars might change the measured intrinsic velocity dispersion. 
We systematically exclude individual stars from the velocity dispersion estimation, one-by-one, and find that star \#2 (denoted in Table \ref{tab:stars}), the largest velocity outlier, causes the largest change by reducing the velocity dispersion to $\sigma_v = 1.9^{+1.4}_{-1.1}$\,\kms. Continuing in this direction, we keep star \#2 out of the member list and systematically exclude individual stars to find the next most impactful source. Removing star \#4, a high S/N measurement at the high-velocity end of the distribution, produces an unresolved velocity dispersion. This systematic analysis of the velocity distribution is relevant because the presence of binary stars in dwarf galaxies can inflate the measured dispersion by several \kms\ relative to the true intrinsic dispersion \citep{McConnachie10, Minor10}, and binary fractions in the ``classical'' dwarf spheroidals has been found to vary broadly, ranging from 14\%\,$-$\,78\% \citep{Minor13, Spencer17, Spencer18, APolonio23}. 

Our spectroscopic measurements indicate that the intrinsic velocity dispersion of UMa3/U1 is resolved, but we note that repeat observations are critical for this to be confirmed. Results from the MCMC calculations for the systematic removal of stars \#2 and \#4 are shown in the central and right-most panels of Figure \ref{fig:mems-mcmc} and identify which sources require particularly careful spectroscopic follow-up observations.

\iffalse
\begin{table}
    \centering
    \begin{tabular}{lcc}
    \toprule
        Stars excluded & $\langle v_{\odot}\rangle$ (km s$^{-1}$) & $\sigma_v$ (km s$^{-1}$) \\
    \midrule
        None & 88.6 & 3.7 \\
        Star \#2 & 89.7 & 1.9 \\
        Stars \#2 \& \#4 & 89.2 & unresolved \\
    \bottomrule
    \end{tabular}
    \caption{Change in the line-of-sight velocity and dispersion as stars are removed that cause the maximum change in the dispersion.}
    \label{tab:disp-rem}
\end{table}
\fi

\subsection{Stellar Mass, Luminosity}\label{subsec:mass}

We now aim to derive the total stellar mass by creating a sample of mock stellar populations that emulate the characteristics of UMa3/U1. We follow a similar methodology to \citet{Martin16-pds} when creating the mock populations.

We assume the underlying stellar population of UMa3/U1 is described by a canonical two-part Kroupa initial mass function (IMF) \citep{Kroupa01} and a stellar population of $\tau = 12$\,Gyr, \met\ $ = -2.2$. We create a single mock stellar population by first drawing a distance from a normal distribution with a mean of 10\,kpc and a standard deviation of 1\,kpc, and shifting the theoretical isochrone to that distance. We then similarly draw a number of stars (N$^*$) from a normal distribution with a mean of 21 and a standard deviation of 5.5 (average of 16\% and 84\% percentiles reported in Table \ref{tab:props}) which will act as the target number of stars above the adopted completeness limit, $i$ = 23.5\,mag. Randomly sampling $D_{\odot}$ and N$^*$ propagate previously derived uncertainties through to the final stellar mass estimation.
We then sample individual stellar masses from the IMF, converting each to the $i$-band and checking if $i_{\text{star}} \geq 23.5$\,mag. We sample the IMF until N$^*$ stars above the completeness limit have been accrued, at which point we sum the stellar mass of all stars (including those below the completeness limit). We repeat this process to create 100,000 mock stellar populations of UMa3/U1 and find the median total stellar mass to be $M_{\text{tot}} = 16^{+6}_{-5}$\,\Msun\ where the uncertainty spans the 16\% and 84\% percentiles of the total stellar mass distribution. This can be recast in terms of the frequentist $p$-value; we reject that the total stellar mass is greater than 38\,\Msun\ at the 99.9\% confidence level.

Additionally, we convert this mass to luminosity and absolute V-band magnitude ($M_V$). We calculate the empirical baryonic mass-to-light ratio (M/L) to be $\sim$\,1.4 for a 12\,Gyr, [Fe/H] = $-2.2$ stellar population, so a stellar mass of $16^{+6}_{-5}$\,\Msun\ implies a total luminosity of 11.4\,$\pm$\,3.6\,\Lsun, which is equivalent to a total absolute $V$-band magnitude of +2.2$^{+0.4}_{-0.3}$\,mag. 
We also compute the effective surface brightness by dividing half the total flux by the area enclosed by one elliptical half-light radius and converting to mag\,arcsec$^{-2}$. This comes to 27\,$\pm$\,1\,mag\,arcsec$^{-2}$. All properties derived here can be found in Table \ref{tab:props}.

Several assumptions are used in this methodology, so we performed several modified analyses to assess the robustness of this result with respect to these choices.
Given a 5$\sigma$ point source depth of 24.0\,mag in $i$, we chose a magnitude cut of $i = 23.5$\,mag to mitigate issues with stellar completeness when deriving stellar parameters, notably the total number of stars in the system. We did not perform a detailed stellar completeness investigation, but we did repeat the same stellar mass analysis using more restrictive magnitude cuts of 23.2\,mag (10$\sigma$ depth) and 22.4\,mag (20$\sigma$ depth). These produced total stellar mass estimates of $21^{+7}_{-6}$\,\Msun\ and $22^{+9}_{-8}$\,\Msun\ respectively, which are consistent with the initial estimate within uncertainties. 
We also repeated the analysis using a Chabrier IMF \citep{Chabrier03}, which produced a nearly identical result of $17^{+6}_{-5}$\,\Msun.
Finally, we investigated the impact of varying age and metallicity as these parameters were constrained by eye alone. We ran the same analysis for isochrones of ages 11 \& 13\,Gyr (holding [Fe/H] fixed at $-2.2$), which gave total stellar masses of 17 and 16\,\Msun\ and absolute $V$-band magnitude was +2.0 and +2.2\,mag, respectively. Similarly, we used isochrones of metallicities $-2.1$ \& $-2.0$ (holding age fixed at 12\,Gyr), producing total stellar masses of 17 and 18\,\Msun\ and absolute $V$-band magnitudes of +2.2 and +2.1\,mag, respectively. For all isochrone changes, we recalculated the empirical baryonic mass-to-light ratio.
This analysis appears to be very robust to small variations in age and metallicity.

%The aim of using a mock stellar population method for calculating magnitude is that by generating 1000s of mocks that stochastically sample a model IMF, 

We note that a common approach to calculate absolute magnitude is to add up the luminosity contribution of every star and obtain an absolute magnitude for each individual mock stellar population \citep[e.g.][]{Martin16-pds, MDelgado22, Collins22, Collins23, McQuinn23-PegW, McQuinn23-LeoMK}. 
%The aim of this methodology is to find a sort of time-averaged representation for the luminosity of a stellar system. While time evolution is not explicitly taken into account, the random sampling of stars at different masses, and therefore different stellar evolutionary phases, should in principle average out to a good representation of the stellar luminosity function. 
This is in contrast to simply counting up the luminosity contribution of each individually observed star on the CMD, which gives the present day luminosity of the system. Direct counting can be challenging due to difficulties in effectively accounting for background/foreground contamination in the member star sample.
UMa3/U1 has a median of 57 total stars (down to to 0.1\,\Msun, over the 100,000 mock populations) meaning that small number statistics play a huge role. Based on the isochrones used to model UMa3/U1, the most massive star is 0.8\,\Msun\ while the MSTO is at $\sim 0.77$\,\Msun. Late-stage stars (RGB, HB) have a massive contribution to the total luminosity of such a tiny system, so mock populations that happen to sample a single star in the mass range $0.77 - 0.8$\,\Msun\ have a dramatically boosted total absolute magnitude, skewing the distribution. 
For this reason, we consider total stellar mass to be a more stable tracer of the stellar content of UMa3/U1, thus providing an estimate of absolute magnitude whose variance is less heavily affected by the occasional sampling of a single late-stage star.

In \citet{Smith22}, we developed a similar method to estimate absolute magnitude of the Bo\"otes V UFD. This methodology was found to produce an excess in the number of bright, late-stage stars (RGB, HB) which resulted in an over-estimation of the total absolute magnitude. We re-derive the total stellar mass and absolute magnitude of Bo\"otes V following the methodology described in this Section by producing 1,000 realisations using stellar population and structural parameters found in \citet{Smith22}. We measure the stellar mass of Bo\"otes V to be $1044^{+775}_{-485}$\,\Msun, which, when converted to absolute $V$-band magnitude gives $M_V = -2.4^{+0.7}_{-0.6}$\,mag. This is consistent with the $V$-band magnitude found by \citet{Cerny23-6} using deeper, targeted follow-up observations from GMOS on Gemini North, measured to be $-3.2^{+0.3}_{-0.3}$\,mag.

\begin{figure}
    \centering
    \includegraphics[width=\linewidth]{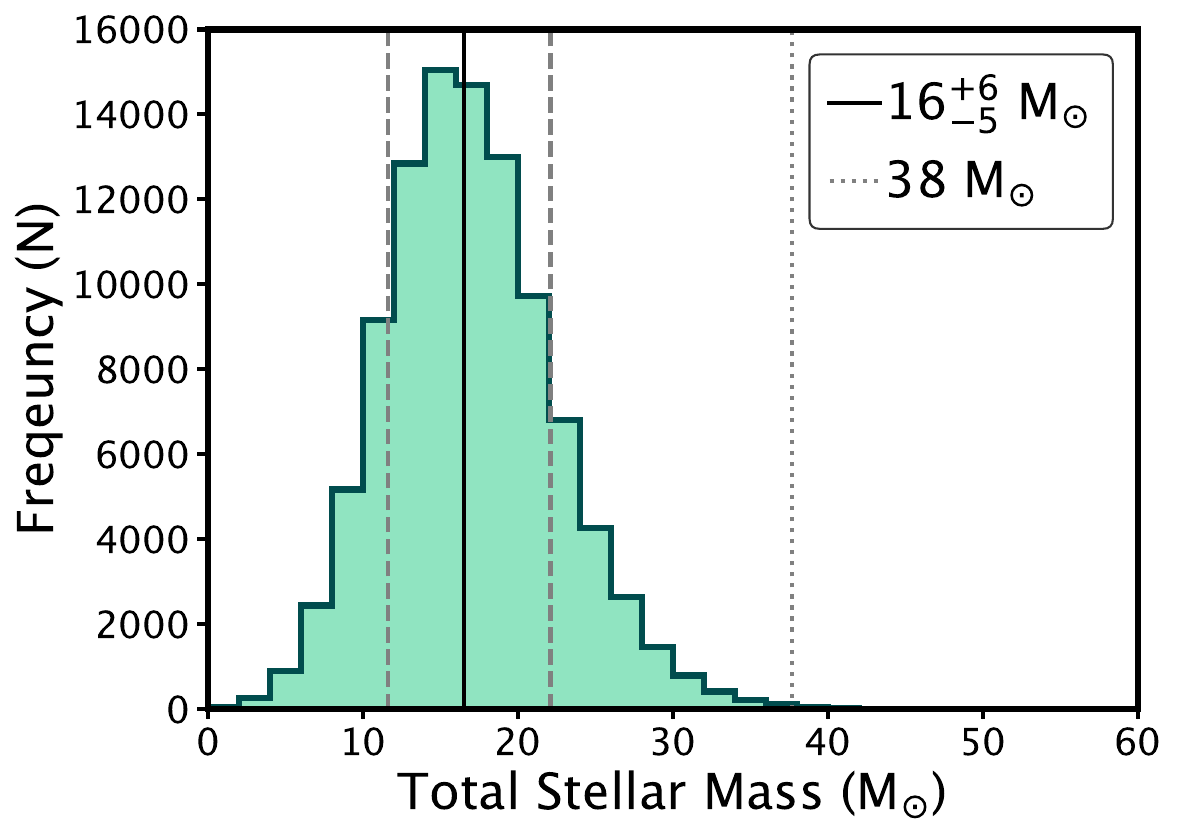}
    \caption{
    Distribution of total stellar mass calculated by creating mock stellar populations of UMa3/U1. The solid black line indicates the median value ($\sim$\,16\,\Msun), the dashed grey lines on either side of the solid line show the 16\% and 84\% percentiles, and the left-most, dotted grey line indicates the 99.9\% confidence level for the upper bound on the stellar mass ($\sim$\,38\,\Msun).
    }
    \label{fig:mass}
\end{figure}

\subsection{Orbital Estimation}\label{subsec:orbit}

\begin{figure*}
    \centering
    \includegraphics[width=\linewidth]{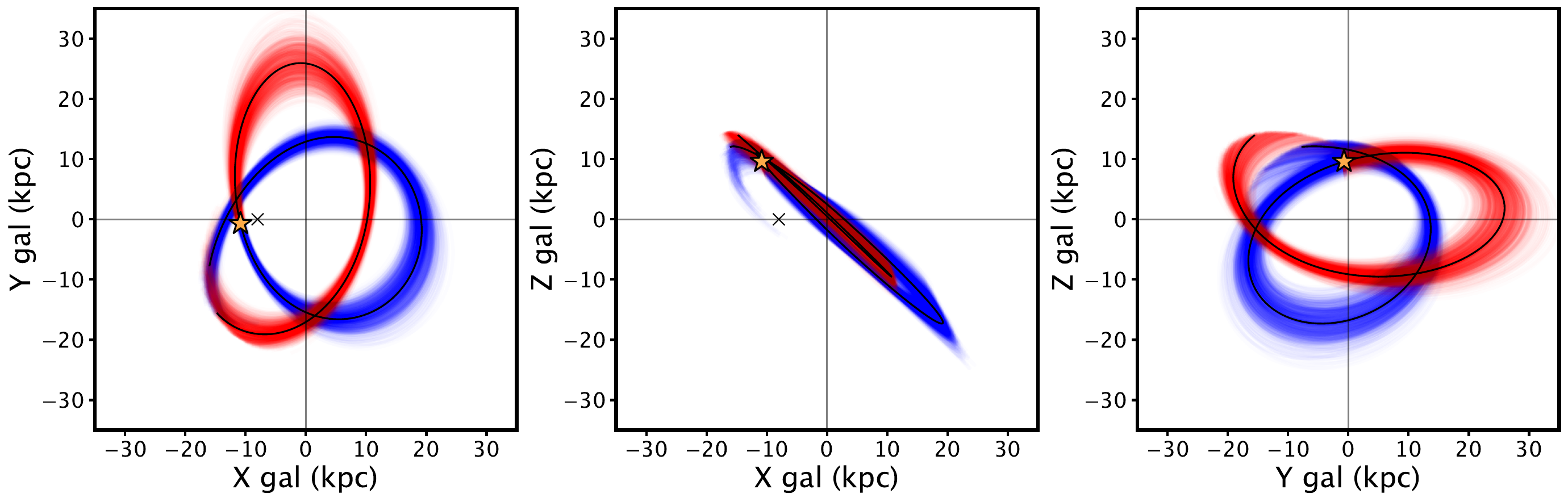}
    \caption{Mean and distribution of orbits resulting from MC analysis in Section \ref{subsec:orbit}, plotted in the galactic XY, XZ, \& ZY planes (from left to right), where the rotation of the Milky Way proceeds clockwise in the XY plane. UMa3/U1 is indicated as a yellow star while the position of the sun on this coordinate system is shown as a black `$\times$'. 
    The orbit is integrated both backwards (blue tracks) and forwards (red tracks) in time by 0.5\,Gyr in steps of 10$^{-3}$\,Gyr from the starting point of each orbit. The mean orbit goes through the yellow star, but each individual orbit has its own starting point, as the position and heliocentric distance are randomized in the MC procedure.}
    \label{fig:orbits}
\end{figure*}

With estimates for all six phase space parameters available, we now use a simple dynamical model to investigate the orbit of UMa3/U1 and its interaction history with the Milky Way. We approximate UMa3/U1 as a point-mass in a Milky Way potential, implemented with the python-wrapped package \textsc{gala} \citep{Price17}. The Milky Way potential used for this analysis is comprised of three components: (1) a \citet{Miyamoto75} disk, (2) a \citet{Hernquist90} bulge, and (3) a spherical NFW dark matter halo \citep{Navarro96}. The parameters used for the bulge and disk are taken from the listed citations whereas the NFW dark matter halo parameters ($M^{\text{DM}}_{\text{200,MW}}$, $R_{200}$) are adopted from estimates by \citet{Cautun20}, where a concentration parameter of 12 is chosen. This potential produces a circular velocity at the radius of the Sun similar to a recent estimate \citep[$v_{circ}(R_{\odot}) = 229$\,\kms;][]{Eilers19}.
We use a right-handed Galactocentric coordinate system such that the Sun is located at (X, Y, Z) = (8.122, 0.0, 0.0) kpc, with local-standard-of-rest velocities of [U,V,W] = [10.79, 11.06, 7.66]\,\kms\,\citep{Robin22}. 

To characterize the orbit, we perform a Monte Carlo randomisation, where we generate 1,000 samples of the initial orbital conditions (i.e. input parameters \{~$\alpha_{J2000}$, $\delta_{J2000}$, $D_{\odot}$, $\mu_{\alpha}$ cos$\delta$, $\mu_{\delta}$, $v_r$~\}), which are previously measured in this analysis. Each parameter, aside from sky positions  ($\alpha_{J2000}$, $\delta_{J2000}$) as uncertainties are negligible, is modelled by a Gaussian distribution with standard deviation given by errorbars indicated in Table \ref{tab:props}. The orbit of each point-mass is integrated 0.5\,Gyr both forwards and backwards in time in steps of $10^{-3}$\,Gyr. Figure \ref{fig:orbits} displays the orbits of all 1,000 realisations, with the blue tracks tracing backwards in time and the red tracks tracing forwards in time. The coordinate system is arranged such that Milky Way rotation proceeds clockwise on the {\it xy}-plane depicted in the left panel of Figure \ref{fig:orbits}, meaning that the orbit of UMa3/U1 is prograde.
The mean orbit is shown in black and the distribution of all 1,000 realisations shows that the orbit is quite stable to uncertainties on input parameters. 
Several key orbital parameters, namely the pericenter ($r_{\text{peri}}$, closest approach to Milky Way), apocenter ($r_{\text{apo}}$, furthest point from Milky Way), $z_{\text{max}}$ (maximum height above the disk), time between pericenters (orbital time), time since last pericenter, and orbital eccentricity, are calculated for each orbit, and the median, along with the 16\% and 84\% percentiles on each parameter, are presented in Table \ref{tab:props}. 

The potential that we use does not include the Large Magellanic Cloud (LMC). We note that \citet{Pace22} computed the change in orbital parameters of all the known UFDs (at the time) given the inclusion and exclusion of the LMC. With $r_{\text{peri}}$, $r_{\text{apo}}$ of 12.9 and 26\,kpc respectively, UMa3/U1 has a smaller apocentre than all the UFDs considered by \citet{Pace22}, and a smaller pericentre than all but one UFD in the \citet{Pace22} list when integrated in a Milky Way-only potential. To find the best sample to compare to UMa3/U1, we selected all UFDs in \citet{Pace22} with $r_{\text{peri}} < 30$\,kpc and $r_{\text{apo}} < 50$\,kpc. This group comprises Tucana III, SEGUE 1, SEGUE 2, and Willman 1. Of these, Tucana III is on a nearly radial orbit and thought to be tidally disrupting. SEGUE 1, SEGUE 2, and Willman 1 are all relatively unaffected by the inclusion of the LMC in the gravitational potential in \citet{Pace22}. Given that UMa3/U1 orbits more closely to the Milky Way than any of these UFDs, we conclude that its orbit is unlikely to be strongly affected by the LMC.

\section{Discussion} \label{sec:disc}

We have presented the discovery of UMa3/U1, an old ($\tau > 11$\,Gyr), metal-poor ([Fe/H] $\sim -2.2$), tiny (3\,$\pm$\,1\,pc) Milky Way satellite with an orbit that remains within $\sim$\,25\,kpc of the galactic center. Most notably, UMa3/U1 is comprised of astonishingly few stars. We have estimated that the total stellar mass is 16$^{+6}_{-5}$\,\Msun, which, when converted to magnitudes using M/L $\sim 1.4$, gives a total absolute $V$-band magnitude of +2.2$^{+0.4}_{-0.3}$\,mag, making UMa3/U1 least luminous Milky Way satellite, and by some margin. Of the faint, ambiguous Milky Way satellites, the faintest are Kim 3 \citep[$M_V$ = +0.7\,mag;][]{Kim16-K3} and DELVE 5 \citep[$M_V$ = +0.4\,mag;][]{Cerny23-6}. Recasting these magnitudes into total stellar mass (again assuming M/L $\sim 1.4$), Kim 3 has $M_{\text{tot}} \sim 63$\,\Msun\ while DELVE 5 has $M_{\text{tot}} \sim 83$\,\Msun. Virgo I \citep{Homma18}, the least luminous presumed dwarf galaxy which is classified based on its physical half-light radius of 47\,pc, has an absolute V-band magnitude of $-0.7$\,mag, representing a total stellar mass of $\sim 230$\,\Msun. All told, UMa3/U1 is a quarter the mass of the previously least luminous Milky Way satellite and some 15$\times$ less massive (in terms of stellar mass) than the faintest suspected dwarf galaxy. We now offer some interpretations regarding the nature of UMa3/U1 as well as the origins of this faint satellite system.

\subsection{On the Origin of Ursa Major III/UNIONS 1}\label{subsec:origin}

\begin{figure*}
    \centering
    \includegraphics[width=\linewidth]{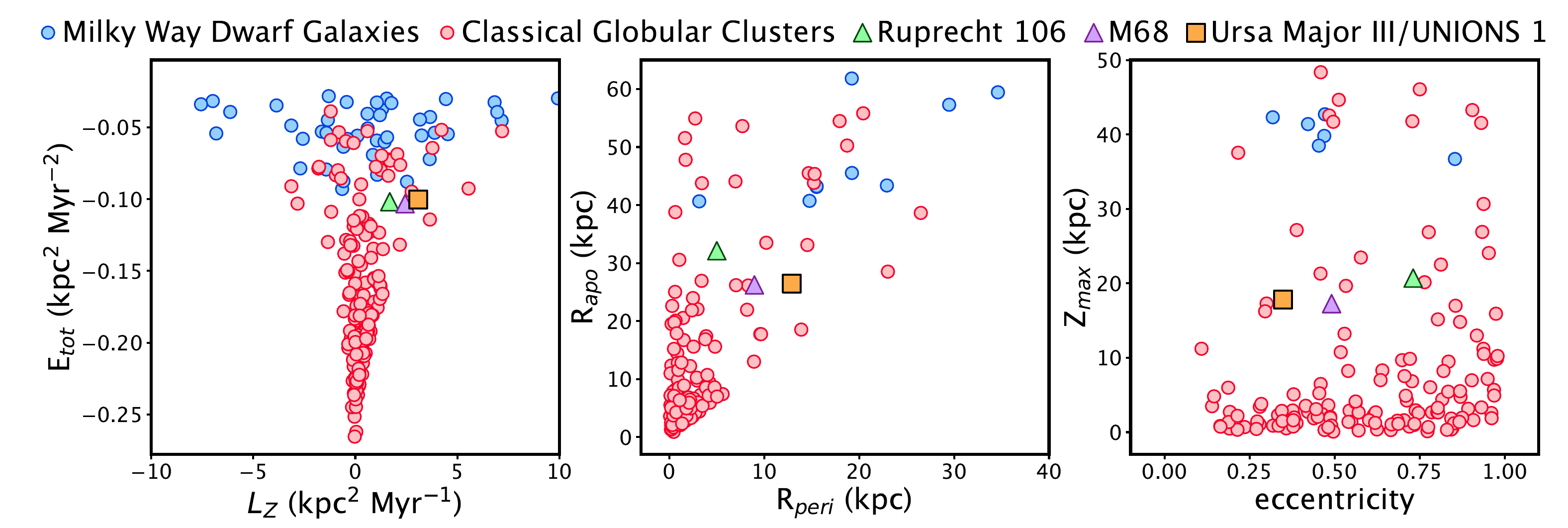}
    \caption{
    {\it Left}: Total energy plotted against the Z-component of the angular momentum for Milky Way globular clusters (red circles), dwarf galaxies (blue circles) and UMa3/U1 (yellow square). The systems with similar orbits to UMa3/U1 are M68 (purple triangle) and Ruprecht 106 (green triangle).
    {\it Center}: Apocenter distance (in kpc) plotted against pericenter distance (in kpc) with the same coloring as in the left panel.
    {\it Right}: Maximum height above/below the Milky Way disk (in kpc) plotted against orbital eccentricity with the same coloring as in the left panel.
    }
    \label{fig:all-orbits}
\end{figure*}

Broadly speaking, there are two possible origins for UMa3/U1: it either formed in situ or it was accreted into the Milky Way. Based on the orbit derived in Section \ref{subsec:orbit} UMa3/U1 does not appear to be a on a disk- or bulge-like orbit. \citet{Massari19} describe bulge GCs as having $r_{\text{apo}} < 3.5$\,kpc and disk GCs $z_{\text{max}} < 5$\,kpc. UMa3/U1 does not satisfy the criteria for either category, with an apocenter of 25.2\,kpc and $z_{\text{max}} = 16.7$\,kpc. 
The findings of \citet{Leaman13} show that in situ, disk-like globular clusters rarely have a mean metallicity less than $-2$ and while \citet{DiMatteo19} argue that a significant portion of the inner Milky Way halo may be comprised of metal-poor (\met\ $<$ $-1$), heated thick-disk stars, they likely extend to a metallicity of $-2$. The metallicity of UMa3/U1 is not strongly constrained, as the PARSEC isochrone database does not extend lower than a metallicity of \met\ $= -2.2$, but it appears metal-poor nonetheless.
UMa3/U1 is on a prograde orbit with a clear, but not too drastic, inclination with respect to the Milky Way plane and therefore could have formed in situ, though it would be fairly anomalous in its orbit, and somewhat anomalous in its metallicity with respect to the known properties of disk globular clusters.

The alternative is that UMa3/U1 could have been accreted into the Milky Way halo. With an orbital period of 373$^{+32}_{-34}$\,Myr, UMa3/U1 has likely had time to complete many pericentric passages of the Milky Way, which may have led to tidal stripping. There is no stellar stream in the \texttt{galstreams}\,\citep[][]{Mateu23} catalog that matches the position or kinematics of UMa3/U1, but it could be fruitful to search for a faint stream along the orbital path given that this satellite has likely been interacting with the outer disk for many orbits.

UMa3/U1 may have been accreted on its own or as a companion to some larger system, so we computed the orbital properties of all Milky Way satellites with measured velocities and proper motions. At the total orbital energy of UMa3/U1, the orbits of globular clusters M68 and Ruprecht 106 are found to have the most similar X, Y, and Z angular momentum, with M68 being a particularly close match in apocenter and the maximum height above/below the disk when examining other orbital parameters. M68 is previously thought to have been accreted into the Milky Way from a satellite galaxy \citep{Yoon02}, so this orbital similarity could indicate that UMa3/U1 and M68 were accreted as part of the same system. If UMa3/U1 is the tidally stripped remains of a dwarf galaxy then it could have hosted M68 prior to accretion, whereas if UMa3/U1 is a star cluster, perhaps it and M68 formed in the same environment. We present the orbital parameters of all Milky Way satellites in Figure \ref{fig:all-orbits} where Milky Way dwarf galaxies are all dwarfs within 420\,kpc ($\sim$ the distance to Leo T) from \citet{McConnachie12}\footnote{\url{https://www.cadc-ccda.hia-iha.nrc-cnrc.gc.ca/en/community/nearby/}}, ``Classical Globular Clusters'' measurements are taken from \citet[][2010 edition]{Harris96}, and the most similar systems, globular clusters M68 and Ruprecht 106, are highlighted.

Although the low metallicity of UMa3/U1 does not exclude it from an in situ formation in the heated thick disk, the orbital parameters are inconsistent with the criteria used to define disk and bulge globular cluster populations. We favor a scenario where UMa3/U1 was accreted into the Milky Way halo.

\subsection{On the Nature of Ursa Major III/UNIONS 1}\label{subsec:nature}

\begin{figure*}
    \centering
    \includegraphics[width=0.75\linewidth]{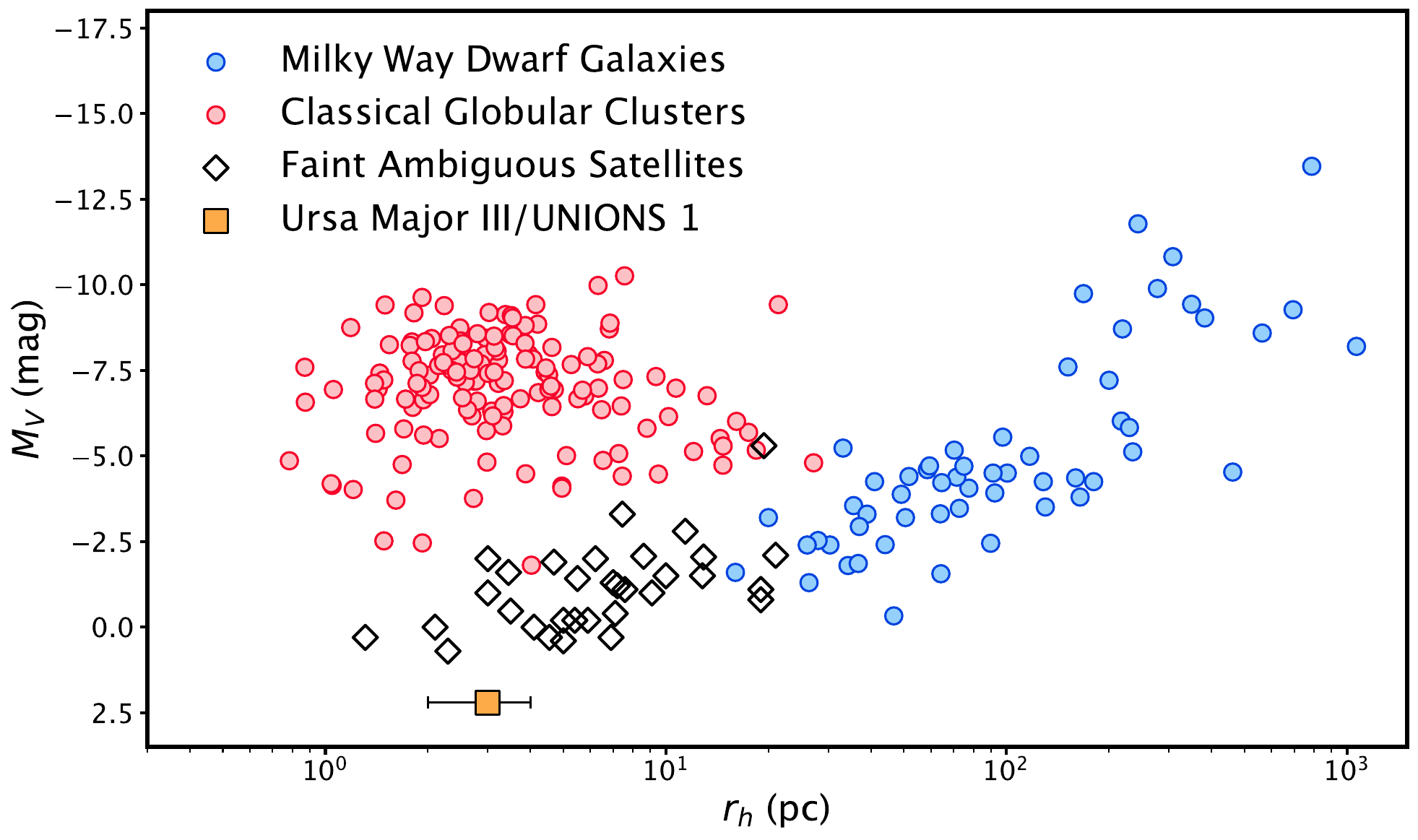}
    \caption{
    $M_V - r_{\text{h}}$ plane with all known Milky Way satellites included. Dwarf galaxies are plotted in blue, ``Classical Globular Clusters'' are plotted in red (where classical refers to those in the Harris Catalog), and faint, ambiguous Milky Way satellites are open black diamonds. UMa3/U1 is shown as as an orange square with \rh\ measurement uncertainties. $M_V$ uncertainties are about the same size as the square marker.
    }
    \label{fig:rhMv}
\end{figure*}

The $M_V - r_{\text{h}}$ plane helps visualize the traits typical of dwarf galaxies, globular clusters, and the faint satellites whose nature remains ambiguous. We have reconstructed this space in Figure \ref{fig:rhMv} where ``Classical Globular Cluster'' and Milky Way dwarf galaxies are taken from the same references as for Figure \ref{fig:all-orbits}, and the faint satellite measurements were compiled from literature. A full list of references can be found in Appendix \ref{ap:sats}. UMa3/U1 is far fainter and smaller than any confirmed Milky Way dwarf galaxies, and lies in a size range occupied by faint, ambiguous satellites and globular clusters.

As suggested by \citet{Willman12}, taken in the context of the $\Lambda$CDM framework, dwarf galaxies reside in their own dark matter halos while globular clusters do not. Dynamical mass estimators \citep[e.g.][]{Wolf10, Errani18} rely on the intrinsic stellar velocity dispersion within a system, which can then be compared with the total stellar luminosity to yield a dynamical mass-to-light ratio. The faintest dwarf galaxies have been measured to have M/L in excess of 10$^3$ \Msun/\Lsun\,\citep[][and references therein]{Simon19} while globular clusters typically have M/L $\sim$ 2\,\citep{Baumgardt20}, consistent with strictly baryonic mass being present. 

If UMa3/U1 is a star cluster, and is therefore composed solely of baryonic matter, then we can use the measured properties of total stellar mass and half-light radius to predict the line-of-sight velocity dispersion of its constituent stars using the mass estimator of \citet{Wolf10}:

\begin{equation}
    \text{M}_{1/2} = 930\cdot\sigma_v^2\cdot r_{\text{h}}\,\text{M$_{\odot}$}
\end{equation}

\noindent where $\sigma_v$ is given in \kms, \rh\ is given in pc, and M$_{1/2}$ is the mass enclosed within \rh. Solving for $\sigma_v$, we estimate $\sigma_v \sim 50$\,m\,s$^{-1}$.

However, in Section \ref{subsec:velocity}, we measured the intrinsic velocity dispersion to be 3.7\,\kms\ from 11 member stars. Now, given $\sigma_v$, we compute the dynamical mass-to-light ratio of UMa3/U1 by dividing Eqn. (6) by the luminosity enclosed within \rh, L$_{1/2}$, which is found by converting $M_{\text{tot}}$ to $L_{\text{tot}}$ using a baryonic M/L $\sim 1.4$ and taking half the result. We carry out the calculation using a 10$^6$-realisation Monte Carlo procedure to propagate measurement uncertainties, where each input quantity is modelled as a Gaussian distribution with a mean and standard deviation given by the values listed in Table \ref{tab:props}. For quantities with uneven uncertainties, we adopt the larger of the two bounds. The dynamical mass-to-light ratio is measured to be 6500$^{+9100}_{-4300}$\,\Msun/\Lsun, implying the presence of a massive dark matter halo, and that UMa3/U1 is a dwarf galaxy with astonishingly little stellar mass.

In Section \ref{subsec:velocity}, we already discussed how the presence of binary stars can inflate the measured dispersion, and we identified the member stars whose velocities and uncertainties appear to contribute most significantly to the estimated value of 3.7\,\kms. The removal of star \#2 from the membership list led to $\sigma_v = 1.9^{+1.4}_{-1.1}$\,\kms, which translates to a dynamical mass-to-light ratio of 1900$^{+4400}_{-1600}$\,\Msun/\Lsun. The volatility of the velocity dispersion with respect to the inclusion of certain candidate member stars makes it unclear as to whether it accurately represents the underlying gravitational potential of UMa3/U1.

%Multi-epoch spectroscopy will be required to conclusively demonstrate the robustness of this dispersion.

Additionally, the use of the \citet{Wolf10} mass estimator assumes that the system being assessed is in dynamical equilibrium. The orbit of UMa3/U1 has a pericenter of 12.8$^{+0.7}_{-0.8}$\,kpc and passes through the disk around 16\,kpc from the galactic center where stellar mass density is $\sim$\,1/50th of that at the solar neighbourhood \citep{Lian22}. It may be the case that repeated interactions with the outer Milky Way disk has led to tidal stripping, which could mean that some of the stars identified as members are in the midst of being stripped and have become unbound. 
If some of the stars that are actively becoming unbound have been observed as part of our spectroscopic follow-ups, their velocities would not be indicative of the gravitational potential underlying UMa3/U1. This could lead to an inflation of the measured velocity dispersion and a subsequent overestimation of the dynamical mass-to-light ratio. 
We investigated the Keck/DEIMOS velocity data by searching for a velocity gradient along the major axis of UMa3/U1, but no clear gradient is visible. We also might expect that stars in the outskirts would be unbound if there is active stripping, which could give them larger velocities relative to the mean. We reran the velocity dispersion MCMC estimate algorithm where we removed the three member stars that are most distant from the centroid (stars \#1, \#4, \& \#7). However, this led to a slight increase in the intrinsic velocity dispersion, giving $\sigma_v = 4.4^{+1.9}_{-1.3}$\,\kms, implying that outer stars are not wholly responsible for the well-resolved velocity dispersion. Using these probes of a velocity gradient and outer stars, we do not find clear signs of unbound stars. 

While the measured velocity dispersion of $\sigma_v = 3.7^{+1.4}_{-1.0}$\,\kms\ may be tracing a massive dark matter halo, we emphasize that the presence of binary stars and unbound stars could impact the interpretation of $\sigma_v$ as an direct indicator of dark matter.
Multi-epoch spectroscopic data taken over a sufficiently long time baseline will be particularly crucial for identifying binary stars and assessing the dark matter content of UMa3/U1.

Focused dynamical modelling of the evolution of UMa3/U1 in the Milky Way halo may provide further clues as to whether this faint, tiny system is a dwarf galaxy or a star cluster. ``Micro-galaxies'' and dwarf galaxies embedded in cuspy dark matter halos are predicted to be rather resilient to tidal disruption \citep{Errani20, Errani22} and the very existence of UMa3/U1 may place constraints on various dark matter models \citep{Errani23}. We refer the reader to work by Errani et al. (in prep.) for a detailed analysis of UMa3/U1 and the implications of its survival in the Milky Way halo.

\section{Summary}

Ursa Major III/UNIONS 1 is the least luminous known satellite of the Milky Way. We identified this satellite as a resolved overdensity of stars consistent with an old, metal-poor isochrone in the deep, wide-field survey, UNIONS. With radial velocities (from Keck/DEIMOS) and proper motions (from {\it Gaia}), we have confirmed that UMa3/U1 is a coherent system.

We have measured an intrinsic velocity dispersion of $3.7^{+1.4}_{-1.0}$\,\kms\ which could be interpreted as the signature of a massive dark matter halo. However, we have demonstrated that the measured line-of-sight velocity dispersion (upon which the presence of dark matter is predicated) is highly sensitive to the inclusion of two stars within the sample of 11 candidate members. It is for this reason that we have referred to this system as UMa3/U1 throughout, with its nature as either a star cluster or a dwarf galaxy remaining ambiguous at this time.

With a half-light radius of 3\,pc, UMa3/U1 occupies a scale-length regime that has typically been assumed to contain star clusters, satellites devoid of dark matter. 
There have only been four moderate resolution spectroscopic studies of these faint, ambiguous systems prior to this one: SEGUE 3 \citep{Fadely11}, Mu\~noz 1 \citep{Munoz12}, Draco 2 \citep{Longeard18}, and Laevens 3 \citep{Longeard19}. All of these programs found inconclusive evidence for the presence or lack of a surrounding dark matter halo, as even \citet{Fadely11} were only able to put an upper bound on the mass-to-light ratio, which favored a baryon-only scenario but could not fully rule out the presence of dark-matter within 1$\sigma$ uncertainties. The study presented in this work highlights the need for further medium-to-high resolution, multi-epoch spectroscopic follow-ups for the whole population of faint, ambiguous satellites. With such sparse stellar populations it remains a technical challenge to observe a sufficient number of stars with sufficient accuracy over a sufficient length of time to confidently measure velocity  dispersions in these faint systems. Dedicated observing time to obtain stellar spectra within these satellites may show that some of these previously assumed star clusters are in fact tiny, faint dwarf galaxies hiding in plain sight.

Each newly found satellite of the Milky Way provides an additional target for investigation and implies that contemporary (e.g. DELVE, UNIONS, DESI Legacy Imaging Surveys) and future (e.g. LSST, the {\it Euclid} space telescope) wide-field, digital, photometric surveys will continue to uncover substructure in the halo of the Milky Way. Population-wide studies of these old, faint, metal-poor systems may provide a unique opportunity to understand the processes of star formation, chemical enrichment, and dynamical interactions, as well as the structure of dark matter, extending previously known relationships to parsec length-scales and tens of solar masses.

\section*{Acknowledgments}

We would like to respectfully acknowledge the L\textschwa\textvbaraccent {k}$^{\rm w}$\textschwa\textipa{\ng}\textschwa n Peoples on whose traditional territory the University of Victoria stands and the Songhees, Esquimalt and $\underline{\text{W}}\acute{\text{S}}$ANE$\acute{\text{C}}$ peoples whose relationships with the land continue to this day.

We would like to thank the anonymous referee whose comments and feedback helped improve the depth and clarity of the manuscript.

As stated in individual acknowledgements below, data collection for this work was conducted at several observing sites atop Maunakea. Therefore, the authors wish first to recognize and acknowledge the very significant cultural role and reverence that the summit of Maunakea has always had within the Native Hawaiian community. We are most fortunate to have the opportunity to conduct observations from this mountain.

This work is based on data obtained as part of the Canada-France Imaging Survey, a CFHT large program of the National Research Council of Canada and the French Centre National de la Recherche Scientifique. Based on observations obtained with MegaPrime/MegaCam, a joint project of CFHT and CEA Saclay, at the Canada-France-Hawaii Telescope (CFHT) which is operated by the National Research Council (NRC) of Canada, the Institut National des Science de l’Univers (INSU) of the Centre National de la Recherche Scientifique (CNRS) of France, and the University of Hawaii. This research used the facilities of the Canadian Astronomy Data Centre operated by the National Research Council of Canada with the support of the Canadian Space Agency. This research is based in part on data collected at Subaru Telescope, which is operated by the National Astronomical Observatory of Japan. 
%We are honored and grateful for the opportunity of observing the Universe from Maunakea, which has the cultural, historical and natural significance in Hawaii. 
Pan-STARRS is a project of the Institute for Astronomy of the University of Hawaii, and is supported by the NASA SSO Near Earth Observation Program under grants 80NSSC18K0971, NNX14AM74G, NNX12AR65G, NNX13AQ47G, NNX08AR22G, YORPD20\_2-0014 and by the State of Hawaii. 

Some of the data presented herein were obtained at Keck Observatory, which is a private 501(c)3 non-profit organization operated as a scientific partnership among the California Institute of Technology, the University of California, and the National Aeronautics and Space Administration. The Observatory was made possible by the generous financial support of the W. M. Keck Foundation.

This research has made use of the Keck Observatory Archive (KOA), which is operated by the W. M. Keck Observatory and the NASA Exoplanet Science Institute (NExScI), under contract with the National Aeronautics and Space Administration.

This work has made use of data from the European Space Agency (ESA) mission {\it Gaia} (\url{https://www.cosmos.esa.int/gaia}), processed by the {\it Gaia} Data Processing and Analysis Consortium (DPAC, \url{https://www.cosmos.esa.int/web/gaia/dpac/consortium}). Funding for the DPAC has been provided by national institutions, in particular the institutions participating in the {\it Gaia} Multilateral Agreement. 


\vspace{5mm}
\facilities{CFHT, PS1, {\it Gaia}, Keck:II}


\software{\texttt{astropy} \citep{Astropy13, Astropy18, Astropy22}, \texttt{emcee} \citep{Foreman13}, \texttt{numpy} \citep{Numpy20}, \texttt{scipy} \citep{Scipy20}}



\appendix

\section{References for Faint Satellites}\label{ap:sats}

Here, we list all references that were compiled when investigating the faint Milky Way satellites whose nature remains ambiguous. At present count, this list includes 32 known systems that exist in the halo of the Milky Way meaning that we exclude the large number of recently discovered globular clusters orbiting in the bulge and disk of the Milky Way, although the majority of these systems are brighter than $M_V \sim -3$\,mag anyways. 
Please now find our list of faint satellite references:
Koposov 1 \citep{Koposov07-fsc}, Koposov 2 \citep{Koposov07-fsc}, SEGUE 3 \citep{Fadely11}, Mu\~noz 1 \citep{Munoz12}, Balbinot 1 \citep{Balbinot13}, Kim 1 \citep{Kim15-K1}, Kim 2 \citep{Kim15-K2}, Crater/Laevens 1 \citep{Laevens15-one, Weisz16}, Laevens 3 \citep{Laevens15-three, Longeard19}, Draco II \citep{Laevens15-three, Longeard18}, Eridanus III \citep{Bechtol15, Koposov15-beasts, Conn18}, Pictor I \citep{Bechtol15, Koposov15-beasts, Jerjen18}, SMASH 1 \citep{Martin16-smash1}, Kim 3 \citep{Kim16-K3}, DES 1 \citep{Luque16, Conn18}, DES J0111-1341 \citep{Luque17}, DES J0225+0304 \citep{Luque17}, DES 3 \citep{Luque18}, DES 4 \citep{Torrealba19-9}, DES 5 \citep{Torrealba19-9}, Gaia 3 \citep{Torrealba19-9}, PS1 1 \citep{Torrealba19-9}, To 1 \citep{Torrealba19-9}, BLISS 1 \citep{Mau19}, HSC 1 \citep{Homma19}, DELVE 1 \citep{Mau20}, DELVE 2 \citep{Cerny21}, YMCA-1 \citep{Gatto21}, DELVE 3 \citep{Cerny23-6}, DELVE 4 \citep{Cerny23-6}, DELVE 5 \citep{Cerny23-6}, DELVE 6 \citep{Cerny23-d6}


\bibliography{ref}{}
\bibliographystyle{aasjournal}



\end{document}